\documentclass[10pt,journal,compsoc]{IEEEtran}

\usepackage{makecell}
\usepackage{hyperref}
\usepackage{array}
\usepackage{graphicx,amssymb,amsmath}
\usepackage{multicol}
\usepackage[noadjust]{cite}
\usepackage{setspace}
\usepackage{subcaption}
\usepackage{graphicx}
\usepackage{float}
\usepackage {url}
\usepackage{stfloats}
\usepackage{amsthm,pifont}
\usepackage{flushend}
\usepackage{cases,subeqnarray}
\usepackage{bm,multirow,bigstrut}
\usepackage{textcomp}
\usepackage{latexsym,bm}
\usepackage{booktabs}
\usepackage{xcolor}
\usepackage{mathtools}
\usepackage{dsfont}
\usepackage{extarrows}
\usepackage{epsfig}
\usepackage{epsfig}
\usepackage{epstopdf}
\usepackage{makecell}
\usepackage{colortbl}
\usepackage{booktabs}
\usepackage{graphicx} 
\usepackage{array}    
\usepackage{hyperref} 
\usepackage{booktabs}
\usepackage{enumitem}
\usepackage{diagbox}
\usepackage{tikz}
\usetikzlibrary{trees,shadows.blur}
\usepackage{forest}
\usepackage{longtable}
\usepackage{booktabs}
\usepackage{amsmath}
\usepackage{graphicx}
\usepackage{xcolor}

\usepackage[ruled]{algorithm2e}
\usepackage{algorithmic} 

\usepackage{ragged2e}

\theoremstyle{plain}

\theoremstyle{plain}

\usepackage{amsmath}

\newcommand{\ignore}[1]{{{\color{yellow} }}}
\definecolor{blue-green}{rgb}{0.0, 0.87, 0.87}
\IEEEoverridecommandlockouts
\begin{document}

\title{
Temporal Spectrum Cartography in Low-Altitude Economy Networks: A Generative AI Framework with Multi-Agent Learning
}
\author{Changyuan Zhao, Ruichen Zhang, Jiacheng Wang, Dusit Niyato,~\IEEEmembership{Fellow,~IEEE}, Geng Sun, Hongyang Du,\\
    Zan Li,~\IEEEmembership{Fellow,~IEEE}, Abbas Jamalipour,~\IEEEmembership{Fellow,~IEEE}, Dong In Kim,~\IEEEmembership{Life Fellow,~IEEE}
    \thanks{C. Zhao is with the College of Computing and Data Science, Nanyang Technological University, Singapore, and CNRS@CREATE, 1 Create Way, 08-01 Create Tower, Singapore 138602 (e-mail: zhao0441@e.ntu.edu.sg).}
    \thanks{R.~Zhang, J.~Wang, and D. Niyato are with the College of Computing and Data Science, Nanyang Technological University, Singapore (e-mail: jiacheng.wang@ntu.edu.sg, ruichen.zhang@ntu.edu.sg, dniyato@ntu.edu.sg).}
        \thanks{G. Sun is with College of Computer Science and Technology, Jilin University, China 130012, (e-mail: sungeng@jlu.edu.cn).}
    \thanks{H. Du is with the Department of Electrical and Electronic Engineering,
University of Hong Kong, Hong Kong (e-mail: duhy@eee.hku.hk).}
    \thanks{Z. Li is with the State Key
Laboratory of Integrated Services Networks, Xidian University, Xian 710071,
China (e-mail: zanli@xidian.edu.cn).}
\thanks{A. Jamalipour is with the School of Electrical and Computer Engineering,
University of Sydney, Australia (e-mail: a.jamalipour@ieee.org).}
    \thanks{D. I. Kim is with the Department of Electrical and Computer Engineering, Sungkyunkwan University, Suwon 16419, South Korea (e-mail:dongin@skku.edu).}
    }

\IEEEtitleabstractindextext{%
\begin{abstract}
\justifying
This paper introduces a two-stage generative AI (GenAI) framework tailored for temporal spectrum cartography in low-altitude economy networks (LAENets). LAENets, characterized by diverse aerial devices such as UAVs, rely heavily on wireless communication technologies while facing challenges, including spectrum congestion and dynamic environmental interference. Traditional spectrum cartography methods have limitations in handling the temporal and spatial complexities inherent to these networks.
Addressing these challenges, the proposed framework first employs a Reconstructive Masked Autoencoder (RecMAE) capable of accurately reconstructing spectrum maps from sparse and temporally varying sensor data using a novel dual-mask mechanism. This approach significantly enhances the precision of reconstructed radio frequency (RF) power maps. In the second stage, the Multi-agent Diffusion Policy (MADP) method integrates diffusion-based reinforcement learning to optimize the trajectories of dynamic UAV sensors. By leveraging temporal-attention encoding, this method effectively manages spatial exploration and exploitation to minimize cumulative reconstruction errors.
Extensive numerical experiments validate that this integrated GenAI framework outperforms traditional interpolation methods and deep learning baselines by achieving 57.35\% and 88.68\% reconstruction error reduction, respectively.
The proposed trajectory planner substantially improves spectrum map accuracy, reconstruction stability, and sensor deployment efficiency in dynamically evolving low-altitude environments.

\end{abstract}
}
\maketitle
\IEEEdisplaynontitleabstractindextext
\IEEEpeerreviewmaketitle

\section{Introduction}\label{intro}

Low-altitude economy networks (LAENets) refer to a growing ecosystem of aerial platforms operating at low altitudes, including unmanned aerial vehicles (UAVs), urban air mobility (UAM) systems, and drone-based services for logistics, agriculture, and surveillance \cite{zheng2025uav}. These networks heavily rely on wireless communication technologies for control, coordination, and data exchange.
However, the dynamic nature of LAENets, combined with varying interference conditions, fluctuating channel states, and heterogeneous traffic demands, leads to significant challenges such as spectrum congestion and inefficient frequency management.
Particularly, spectrum information supports a wide range of downstream wireless applications, such as indoor positioning, interference-free routing, and adaptive networking \cite{chen2025dynamic}, all of which are critical for enabling low-altitude economic activities.
Therefore, accurately estimating radio frequency (RF) spectrum usage in LAENets is essential for improving the efficiency and reliability of such social activities.



Spectrum cartography is a technique that enables the visualization and spatial mapping of RF spectrum usage across different geographic locations. It provides detailed insights into signal power, interference power, and power spectral density (PSD) by constructing high-resolution maps of wireless signal propagation across a geographic area \cite{chen2017efficient}.
Conventional spectrum cartography relies on deployed sensors to collect radio information, which is then used to reconstruct a map of the entire area. 
Formally, spectrum mapping is formulated as a sparse inverse problem, where the objective is to recover missing values from limited observations. The primary challenge arises from data sensitivity, emphasizing how sensor placement and the number of sensing points significantly impact the accuracy of the final spectrum mapping.
Numerical methods primarily exploit the spatial smoothness of power propagation as a physical assumption and implement various interpolation techniques. For instance, G. Boccolini et al. utilized Kriging \cite{boccolini2012wireless}, while S. {\"U}reten et al. applied thin-plate splines (TPS) \cite{ureten2012comparison} for spectrum cartography.
With the advancement of artificial intelligence (AI) technology, particularly generative AI (GenAI), which can learn data distributions and supplement missing information based on limited data, an increasing number of spectrum mapping algorithms based on machine learning and deep learning have been proposed. 
These algorithms typically do not rely on physical assumptions, giving them enhanced generalization capabilities.
For example, 
S. Shrestha et al. and Y. Teganya et al. introduced the autoencoder (AE) structure to complete the missing spectrum~\cite{shrestha2022deep, teganya2021deep}. 

The preceding approaches mainly emphasize the deployment and measurement of static sensors, leading to the construction of time-invariant spectrum maps. However, in applications of LAENets, such as air traffic management and logistics, spectrum signals in the environment exhibit temporal variations. Relying solely on static spectrum diagrams poses challenges in accurately supporting operations in these dynamic scenarios \cite{reddy2022spectrum}.
Moreover, with UAVs evolving into mobile wireless sensing and relay units, the focus can shift from static sensor deployment to adaptive sensing strategies, enabling dynamic selection of target areas for more effective spectrum monitoring \cite{shrestha2022spectrum}.
Meanwhile, the complexity of the low-altitude environment, characterized by variable wind speeds, increases uncertainty for sensing equipment, potentially disrupting system decision-making \cite{zhao2025generative}. 
Therefore, developing an effective cartography algorithm is essential.

Recently, the Masked Autoencoder (MAE), a GenAI model that employs a masking mechanism to enhance feature extraction in self-supervised learning, has been widely adopted as the backbone in Transformer and Vision Transformer (ViT) training \cite{tong2022videomae}. MAE enables effective feature representation for data recovery through pre-training and facilitates the learning process of various downstream tasks followed by fine-tuning. This approach has been successfully applied in the development of Large Language Models (LLMs) and Large Vision Models (LVMs) \cite{han2024efficient}.
In addition, diffusion-based reinforcement learning \cite{du2024enhancing}, which formulates policy generation as a denoising process, shifts the focus from deterministic policy modeling to policy distribution. This method has demonstrated promising performance in handling uncertainty and ensuring robust decision-making in dynamic environments \cite{zhao2025generative}.


Based on these achievements of GenAI models, we propose a two-stage GenAI framework, which integrates temporal spectrum cartography with trajectory optimization for dynamic sensors, to tackle the aforementioned challenges.
This framework estimates the power map over time by leveraging both static and dynamic UAV sensor data.
In the first stage, we employ an MAE \cite{he2022masked}, which can reconstruct data from limited input to generate a temporal power map for each time slot. Subsequently, in the second stage, we apply diffusion-based reinforcement learning \cite{du2024enhancing} to determine the optimal placement of dynamic UAV sensors in the next time slot, thereby enhancing spectrum mapping accuracy.
Our main contributions are summarized as follows:
\begin{itemize}
    \item 
    We propose a two-stage GenAI framework 
    that reconstructs the spectrum map
    and optimizes the trajectories of multi-agent dynamic sensors to map the temporal spectrum during a certain period accurately.
    \textit{To the best of our knowledge, this is the first work that 
    considers temporal spectrum cartography with trajectory optimization in dynamic environments.
    } 

    \item 
    We present a Reconstructive Masked Autoencoder (RecMAE) designed to recover the temporal spectrum map through a dual-mask mechanism. The first mask operates at the pixel level, applied directly to the original image to simulate sparsely collected sensor data. The second mask, applied at the patch level, is introduced after patch embedding. Leveraging self-supervised learning, 
    the dual-mask mechanism enhances the model’s ability to learn fine-grained details and reconstruct the spectrum precisely.

    \item 
    For trajectory optimization, we propose a Multi-agent Diffusion Policy (MADP) framework that integrates diffusion-based policy learning into multi-agent learning. We design a temporal-attention state encoder that enables the selection of optimal strategies based on temporal and high-dimensional state information. The framework effectively focuses on critical features by tightly integrating diffusion policies with the attention mechanism, even when only sparse or partial sensory data is available.

    
\end{itemize}

The structure of this paper is as follows. Section \ref{sec:sec2} provides an overview of related research. The system models are introduced in Section \ref{sec:sec3}. Section \ref{sec:sec4} focuses on the detailed design of the proposed two-stage GenAI framework, including RecMAE and MADP. Then, Section \ref{sec:sec5} presents the numerical results, followed by the conclusion in Section \ref{sec:sec6}.

\section{Related Work}\label{sec:sec2}

\subsection{Spectrum Cartography Approaches}

Spectrum cartography aims to characterize and visualize the utilization of RF power over a geographical region. 
Traditional spectrum cartography relies on Maxwell’s equations to model RF signal propagation. However, due to limited computational resources, early methods were restricted to simple scenarios, such as estimating fields from a dipole source \cite{romero2022radio}.
More recent methods adopt spatial interpolation techniques based on assumptions, including path-loss models and signal smoothness. For instance, Kriging-based approaches estimate missing values by exploiting spatial correlations through covariance functions \cite{boccolini2012wireless}, while TPS have shown strong performance with dense sampling \cite{ureten2012comparison}. 
Further research has explored the multidimensional correlation of spectrum data using tensor and matrix completion techniques \cite{tang2016joint, schaufele2019tensor, zhang2020spectrum}. 
In \cite{tang2016joint}, the authors employed coupled block-term decompositions to reconstruct radio maps across varied sensing patterns, ensuring identifiability under systematic sampling. 
Another approach enhanced coverage estimates by combining low-rank and smoothness constraints, using total variation regularization to robustly recover incomplete spatio-spectral data \cite{schaufele2019tensor}. The authors in \cite{zhang2020spectrum} introduced a joint completion-prediction framework that leverages historical measurements to improve reconstruction accuracy.
However, these methods often rely on stationary models, which can struggle in dynamic environments or under sparse sensor deployments.
This limits their ability to adapt to rapidly fluctuating interference and traffic demands, which commonly arise in LAENets.

\subsection{Learning Approaches for Cartography}

With the advancement of machine learning and deep learning technologies, an increasing number of learning-based methods have been proposed for radio map reconstruction and related tasks \cite{han2020two, levie2021radiounet, roger2023deep, shrestha2022deep, teganya2021deep}. Compared to traditional numerical approaches, learning-based algorithms can autonomously extract spectral features via neural networks, thereby reducing reliance on strong prior assumptions.
For instance, in \cite{han2020two}, the authors employed a neural network model trained on sampled data, leveraging transfer learning to adapt to varying deployment scenarios. In \cite{levie2021radiounet}, a convolutional neural network (CNN) was designed to reconstruct radio maps, while \cite{roger2023deep} utilized a long short-term memory (LSTM) network to capture temporal dynamics in V2X communications. Furthermore, \cite{shrestha2022deep} proposed integrating deep learning with nonnegative matrix factorization to enhance reconstruction performance.

Although the above methods adopt different model architectures, most are built upon the AE framework \cite{teganya2021deep}. AE-based generative models offer certain advantages in data generation and recovery due to their ability to learn compact representations of input data. However, the compression inherent to AE structures often leads to information loss, making fine-grained reconstruction difficult \cite{zhao2024generative}.
To address this limitation, advanced generative AI models such as the MAEs~\cite{he2022masked} have been developed.
Unlike conventional AEs, MAEs maintain spatial dimensions, preserving structural details and reducing issues caused by compression.
Hence, the MAE architectures have been adopted widely in ViTs \cite{han2022survey}, LLMs \cite{zhou2024large}, and LVMs \cite{han2024efficient}, enabling powerful generation and completion capabilities across various modalities.
In contrast to existing AE-based methods, this paper proposes adopting an MAE framework to preserve spatial fidelity better and enhance generative reconstruction performance.

\subsection{Optimization Techniques in Cartography}

When transitioning from static to temporal cartography, the design of sensor deployment and mobility strategies becomes crucial. Early approaches typically assumed that sensor nodes were either strictly stationary or deployed based on a one-time optimization procedure \cite{sorour2014joint}.
With the advancement of low-altitude technologies, UAVs have emerged as mobile wireless devices capable of dynamic sensing and communication \cite{gong20233d,zhao2023online,li2023uav}.
For instance, the authors in \cite{gong20233d} studied trajectory planning in three-dimensional space, emphasizing the challenge of maintaining reliable connectivity in aerial space. 
To address connectivity issues, reinforcement learning has been increasingly employed. 
In \cite{zhao2023online}, the authors proposed a reinforcement learning-based trajectory optimization method using an outage probability map to reflect the connection quality between UAVs and ground base stations. 
Similarly, in \cite{li2023uav}, the authors tackled the challenge of sparse and delayed feedback, which hinders the performance of traditional methods. They proposed a Proximal Policy Optimization (PPO)-based framework to learn navigation policies using delayed interaction feedback from base stations, enabling effective decision-making despite limited real-time information.
Despite these advances, most existing works focus on static spectrum cartography and often overlook the potential of cooperation among multiple UAVs, which is the key aspect addressed in this paper.

Furthermore, diffusion policy has emerged as a promising GenAI framework for sequential decision-making \cite{chi2023diffusion}. By generating actions through multi-step denoising processes, diffusion models have demonstrated strong potential in wireless communication tasks such as sensor placement \cite{zhao2025generative2}, resource allocation \cite{zhao2025enhancing}, signal direction of arrival estimation \cite{wang2024generative2}, and beamforming \cite{wang2024generative}. Due to their superior performance in dynamic environments, diffusion policies show particular promise for application in LAENets.
While existing literature has developed powerful techniques for static cartography, applying these methods to dynamic LAENet scenarios remains an open challenge. The convergence of multi-agent learning and GenAI offers a promising research direction to address current limitations. 
To this end, this paper proposes an MADP framework aimed at accurate and scalable temporal spectrum cartography through collaborative UAV sensing and navigation.

\section{System Model}\label{sec:sec3}

In this section, we first introduce temporal spectrum cartography within LAENet scenarios. Then, we formulate temporal radio map cartography as a dynamic tensor completion optimization.

\begin{figure}[htbp]
    \centering
    \includegraphics[width= 0.98\linewidth]{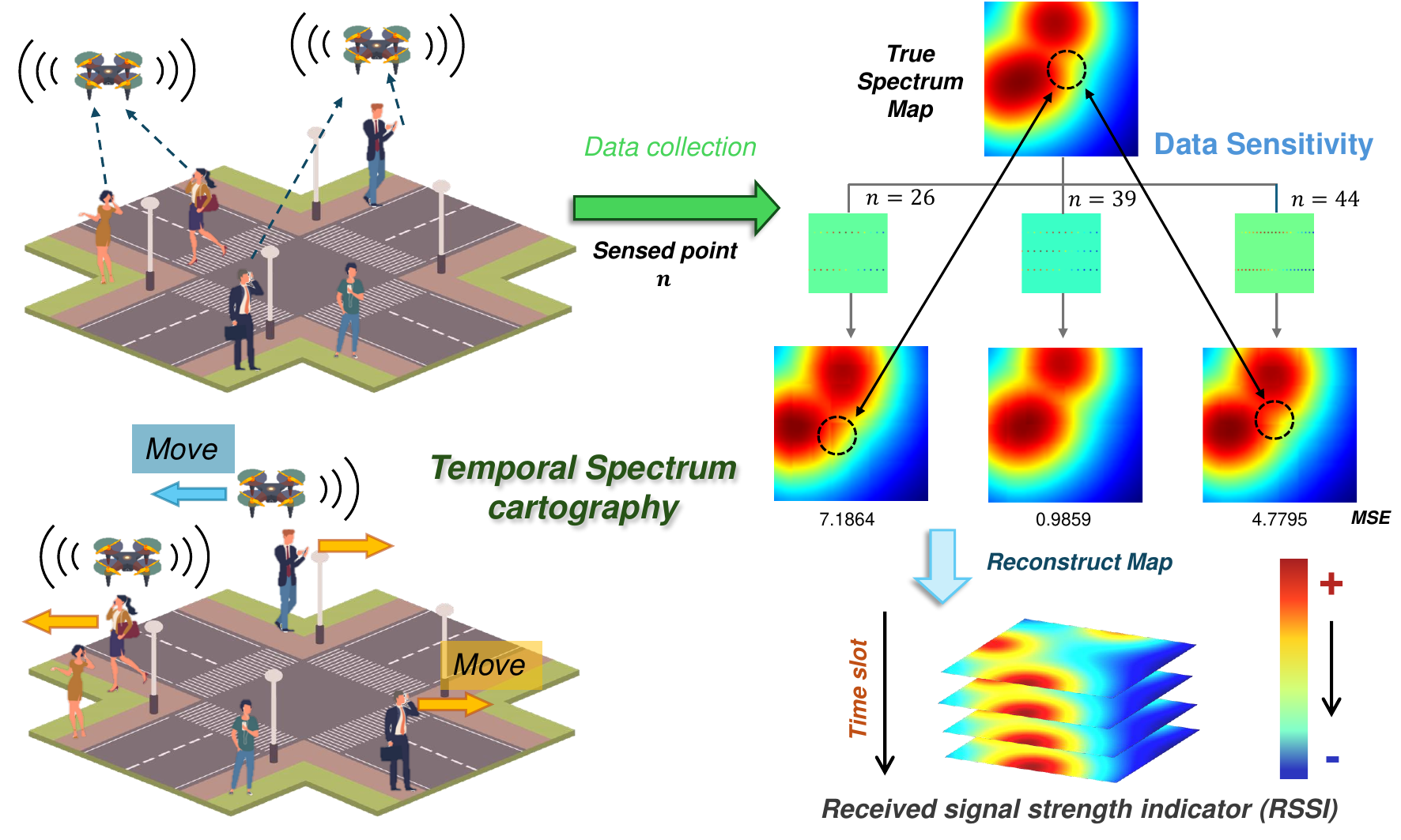}
    \caption{
    System model of temporal spectrum cartography in LAENets. Low-altitude devices (e.g., UAVs) collect RSSI measurements and reconstruct radio maps over time. The reconstruction quality is sensitive to the locations and number of the sensed points.
    }
    \label{fig:sysmod}
\end{figure}

\subsection{System Overview}





As illustrated in Fig. \ref{fig:sysmod}, we consider low-altitude economic activity scenarios in urban environments, where $N$ mobile ground devices engage in communication. Within a predefined range $X$, the propagated signals will form a spectrum diagram.
At a certain height $h_{sensor}$ in the air, low-altitude sensing equipment, consisting of $M$ sensors, which include $M_d$ dynamic UAVs and $M_s$ static sensors, senses the spectrum by receiving these signals. The sensing regions for dynamic drones and static sensors are denoted as $R_d$ and $R_s$, respectively.
Note that in the considered temporal spectrum cartography task, the low-altitude sensing equipment operates at a constant altitude 
$h_{sensor}$ to construct 2-D spectrum maps
\cite{shrestha2022spectrum}.

Given the continuous economic activities in LAENets, monitoring the spectrum in a time period $T_s$ is crucial. We divide the sensing period $T_s$ into $n_T$ time slots $T_i$, i.e., $T_s = \sum_{i=1}^{n_T} T_i$. During each slot, $T_i$, the low-altitude sensing devices remain stationary to ensure stable measurements. To accurately capture spectral variations caused by mobile devices, we assume that data collection occurs at $n_{T_i}$ moments within slot $T_{i}$, where $i=1,\ldots,n_T$.
At the end of each slot, sensor data is aggregated to generate the spectrum for that slot. Subsequently, dynamic sensors adjust their positions for the next slot based on control center instructions to enhance cartography accuracy \cite{yapar2023real}. 
To facilitate analysis, we discretize the sensing area $X$ into a grid with intervals $\Delta x$ and $\Delta y$, forming an $H \times W$ grid. The corresponding gridded spectrum map is then represented as $\mathbf{P} \in \mathbb{R}^{H \times W}$ \cite{romero2022radio}.

\begin{figure*}[htbp]
    \centering
    \includegraphics[width= 0.90\linewidth]{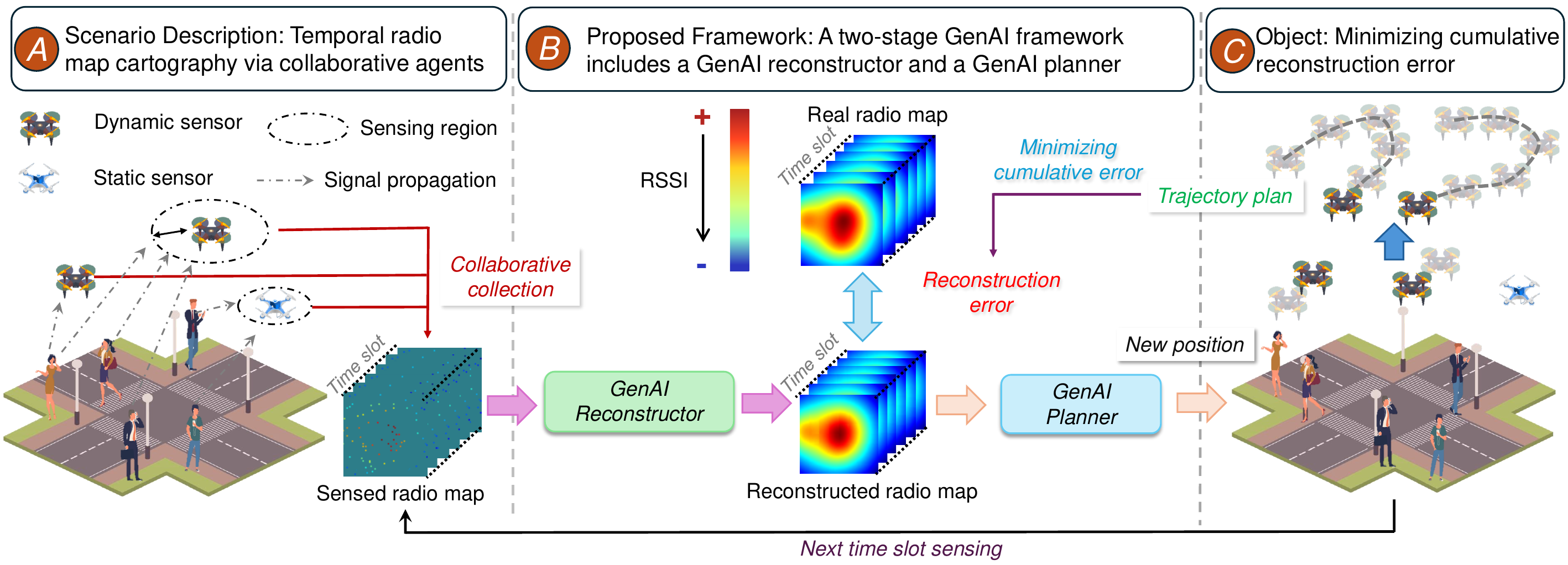}
    \caption{
The workflow of the proposed two-stage GenAI framework.
\textit{Part A} presents the LAENet-based radio map cartography scenario, where dynamic and static sensors collaboratively collect RSSI within a certain sensing region.
\textit{Part B} details the two-stage GenAI framework, consisting of the GenAI-based reconstructor that generates the temporal radio map using sensed data and the GenAI-based planner that determines the next position of dynamic sensors.
\textit{Part C} highlights the objective of this framework: to minimize the cumulative reconstruction error across time slots by optimizing the trajectories of multiple agent sensors, thereby improving the overall quality of the estimated radio map.
}
\label{fig:frame}
\end{figure*}

\subsection{Signal Model}

In this paper, we estimate the power map for spectrum cartography \cite{romero2022radio}. The power can be computed effectively by the received signal strength indicator (RSSI) \cite{xue2018new}.

Specifically, 
we define the set of user equipment (UE) positions as 
\[
\mathcal{T} = \{(x_i, y_i)\}_{i=1}^{N},
\]
where $N$ is the number of UEs, and $\text{UE}_i$ is located at $(x_i, y_i)$.
For a grid point $(x,y)\in X$ and $\text{UE}_i$, the 3D distance between $\text{UE}_i$ and the point with sensors' height $h_{\text{sensor}}$ is given by:
\[
d_{3D} = \sqrt{
d_{2D}^2 + h_{sensor}^2},
\]
where $d_{2D} = \sqrt{(x - x_{i})^2 + (y - y_{i})^2}$ denotes the plane distance.
Additionally, the elevation angle \( \theta \) is given by:
\begin{equation}
    \theta = \arctan\left( \frac{{h_{\text{sensor}}}}{d_{2D}} \right).
\end{equation}
In our study, we consider a probabilistic line-of-sight (LOS) model, which accounts for both LOS and non-line-of-sight (NLOS) conditions.
The LOS probability is computed as \cite{zhu20213gpp}:
\begin{equation}
\label{eq:Los}
    P_{\text{LOS}} = \frac{1}{1 + a_{\text{LOS}} \cdot e^{-b_{\text{LOS}}(\theta - a_{\text{LOS}})}},
\end{equation}
where \( a_{\text{LOS}}\) and \( b_{\text{LOS}}\) are model parameters. Then, the NLOS probability can be computed by:
\begin{equation}
    P_{\text{NLOS}} = 1 - P_{\text{LOS}}.
\end{equation}
The path loss in dB for LOS and NLOS cases is given by~\cite{zhu20213gpp}:
\begin{equation}
    PL_{\text{LOS}} = PL_{d_0} + 10 n_{\text{LOS}} \log_{10} \left(\frac{d_{\text{3D}}}{d_0} \right),
\end{equation}
\begin{equation}
    PL_{\text{NLOS}} = PL_{d_0} + 10 n_{\text{NLOS}} \log_{10} \left(\frac{d_{\text{3D}}}{d_0} \right)+X_{\text{NLOS}},
\end{equation}
where \( PL_{d_0} \) is the reference path loss, \( n_{\text{LOS}}\) and \( n_{\text{NLOS}} \) are path loss exponents, 
$X_{\text{NLOS}} \sim \mathcal{N}(0, \sigma^2_{\text{NLOS}})$ represents a normally distributed random variable modeling shadow fading, and $\sigma_{\text{NLOS}}$ denotes the standard deviation.

Note that the random shadow fading is spatially correlated. For grid points \((x, y)\) and \((m, n)\), it holds that
\begin{equation}
    \mathrm{Cov}[X_{\text{NLOS}}^{x,y}, X_{\text{NLOS}}^{m,n}] = \sigma_{i}^2 e^{-\frac{d((x,y),(m,n))}{d_{\text{corr}}}},
\end{equation}
where \(\mathrm{Cov}[\cdot,\cdot]\) denotes the covariance, 
\( d(\cdot,\cdot) \) denotes the distance between two grid points, and \( d_{\text{corr}} \) is the shadowing decorrelation distance~\cite{gudmundson1991correlation}.
We focus on the signal strength at a specific frequency $f$ to support targeted low-altitude economic activities. Additionally, we consider low-speed scenarios where the frequency shift caused by the Doppler effect can be neglected.
Based on these assumptions,
the reference path loss can be computed by:
\[
PL_{d_0} = 20 \log_{10}(\lambda / (4 \pi d_0)), 
\]
where $d_0$ is the reference distance, $\lambda=c/f$ is the wavelength, $c$ denotes the speed of light, and $f$ represents the sensed frequency.
Then, the received power from $\text{UE}_i$ at grid point $(x,y)$ can be defined as follows:
\begin{equation}
\label{eq:LOSNLOS}
    P_{x,y,i}^{\text{dBm}} = P_{\text{UE}_i}^{\text{dBm}} - \left( P_{\text{LOS}} PL_{\text{LOS}} + P_{\text{NLOS}} PL_{\text{NLOS}} \right),
\end{equation}
where $P_{\text{UE}_i}^{\text{dBm}}$ represents the transmission power of $\text{UE}_i$.
Finally, the power map is updated as:
\begin{equation}
    P_{x,y} = \sum_{i=1}^{N}
    10^{P_{x,y,i}^{\text{dBm}} / 10},
\end{equation}
which can be converted back to dBm by
\begin{equation}
\label{eq:generateDBM}
    P_{x,y}^{\text{dBm}} = 10 \log_{10} P_{x,y},
\end{equation}
where $P_{x,y}^{\text{dBm}}$ is the $(x,y)$-th element of radio power map $\mathbf{P}$.

\subsection{Temporal Tensor Completion}

Following the signal model, we define the power map at time slot $T_i$ as
$\mathbf{P}^{i}\in\mathbb{R}^{n_{T_i}\times H \times W}$.
The sensing operation at the time slot $T_i$ is represented by the binary matrix
$\mathbf{W}^{i} \in\mathbb{R}^{H\times W}$, where 
each element $\mathbf{W}^{i}_{x,y} = 1$ if and only if $(x,y)\in X$ falls within the sensing range of the deployed sensors, as illustrated in Fig. \ref{fig:frame} \textit{Part A}.
Consequently, the sensed power map can be expressed as follows:
\begin{equation}
    \mathbf{\tilde{P}}^i = \mathbf{W}^{i}\circ\mathbf{P}^{i},
\end{equation}
where $\circ$ represents the Hadamard (element-wise) product.
We define the reconstructed power map as $\mathbf{\hat{P}}^i$ to recover the complete power map.
The reconstruction error of time slot $T_i$ is then given by
\begin{equation}
    E^{i} = \vert\vert \mathbf{\tilde{P}}^i - \mathbf{\hat{P}}^i \vert\vert_2.
\end{equation}
By aggregating the reconstruction errors across all time slots, the total reconstruction error over the period $T$ is
\begin{equation}
    E = \sum_{i=1}^{n_T} E^{i}.
\end{equation}

The reconstruction quality heavily depends on the sensor placement and the number of deployed sensors.
As shown in Fig. \ref{fig:sysmod}, improper placement can significantly degrade the algorithm's performance even with a larger number of sensors.
Therefore, optimizing the sensor placement at each time slot, which is dictated by $\mathbf{W}^{i}$, to minimize the total reconstruction error constitutes a temporal tensor completion problem. Unlike static tensor completion, where the observed entries are fixed, dynamic tensor completion aims to adaptively select the observed location over time to enhance reconstruction quality.
Mathematically, the goal is to optimize $\mathbf{W}^{i}$ such that
\begin{equation}
    \min_{\mathbf{W}^1, \dots, \mathbf{W}^{n_T}} E = \sum_{t=1}^{n_T} \left\| \tilde{\mathbf{P}}^t - \hat{\mathbf{P}}^t \right\|_2,
\end{equation}
as shown in Fig. \ref{fig:frame} \textit{Part C}.

Let $\mathcal{U}_t$ denote the positions of the dynamic UAV sensors at time slot $t$, and let $\mathcal{S}$ denote the set of static sensors.
We consider that each dynamic UAV sensor can move with a limited distance between consecutive time slots. For simplicity, we model this constraint as a movement between adjacent grids, denoted by $d_m$.
Formally, the optimization problem is formulated as follows:
\begin{align}
\min_{\mathbf{W}^1, \dots, \mathbf{W}^{n_T}} \quad & E \\
\label{cs:1}
\text{s.t.} \quad & \mathbf{W}^t_{i,j} = 1 \iff \notag \\
&\left( 
\begin{array}{l}
|d\big((x,y), (x^t_u, y^t_u)\big)| \leq R_d \quad \text{or} \\
|d\big((x,y), d(x_s, y_s)\big)| \leq R_s
\end{array}
\right),\notag  \\
& \hspace{1em} \forall (x^t_u, y^t_u) \in \mathcal{U}_t,\ (x_s, y_s) \in \mathcal{S},\ t = 1, \dots, n_T, \\[1ex]
\label{cs:2}
& |d\big((x^t_u, y^t_u), d(x^{t+1}_u, y^{t+1}_u)\big)| \leq d_m, \notag \\
& \hspace{2em} \forall (x^t_u, y^t_u) \in \mathcal{U}_t,\ (x^{t+1}_u, y^{t+1}_u) \in \mathcal{U}_{t+1}, \\[1ex]
\label{cs:3}
& |\mathcal{U}_t| = M_d,\quad |\mathcal{S}| = M_s,\quad \forall t = 1, \dots, n_T. 
\end{align}
The constraint in~\eqref{cs:1} ensures that the element \( \mathbf{W}^t_{i,j} \) is sensed by either dynamic or static sensors.  
The constraint in~\eqref{cs:2} specifies the movement range of dynamic UAVs during each time slot.  
The constraint in~\eqref{cs:3} ensures the number of sensors remains constant during the sensing period.

\section{Proposed Two-stage Generative AI Framework}\label{sec:sec4}

In response to the aforementioned optimization, we propose a two-stage GenAI framework, as illustrated in Fig. \ref{fig:frame} \textit{Part B}.
First, we introduce a GenAI reconstructor that reconstructs the radio power map from sensed data. The proposed GenAI reconstructor leverages the generative capability to achieve higher reconstruction accuracy than existing methods.
Next, we design a GenAI planner to optimize UAV positions using multi-agent reinforcement learning. By utilizing its distribution learning ability, GenAI can generate a more reliable multi-agent policy, effectively balancing exploration and exploitation.
In this section, we define the proposed framework and its key components.

\subsection{Reconstructive Masked Autoencoder}

For the GenAI reconstructor, we propose a RecMAE model that extends the MAE \cite{he2022masked} for recovering spatio-temporal radio map sequences.
In the following, we detail the key components and formulation of RecMAE, including the attention mechanism, patch embedding, positional encoding in both space and time, the encoder-decoder structure, temporal modeling across frames, dual-mask mechanism, and the loss functions used for training.

\begin{figure*}[htbp]
    \centering
    \includegraphics[width= 0.85\linewidth]{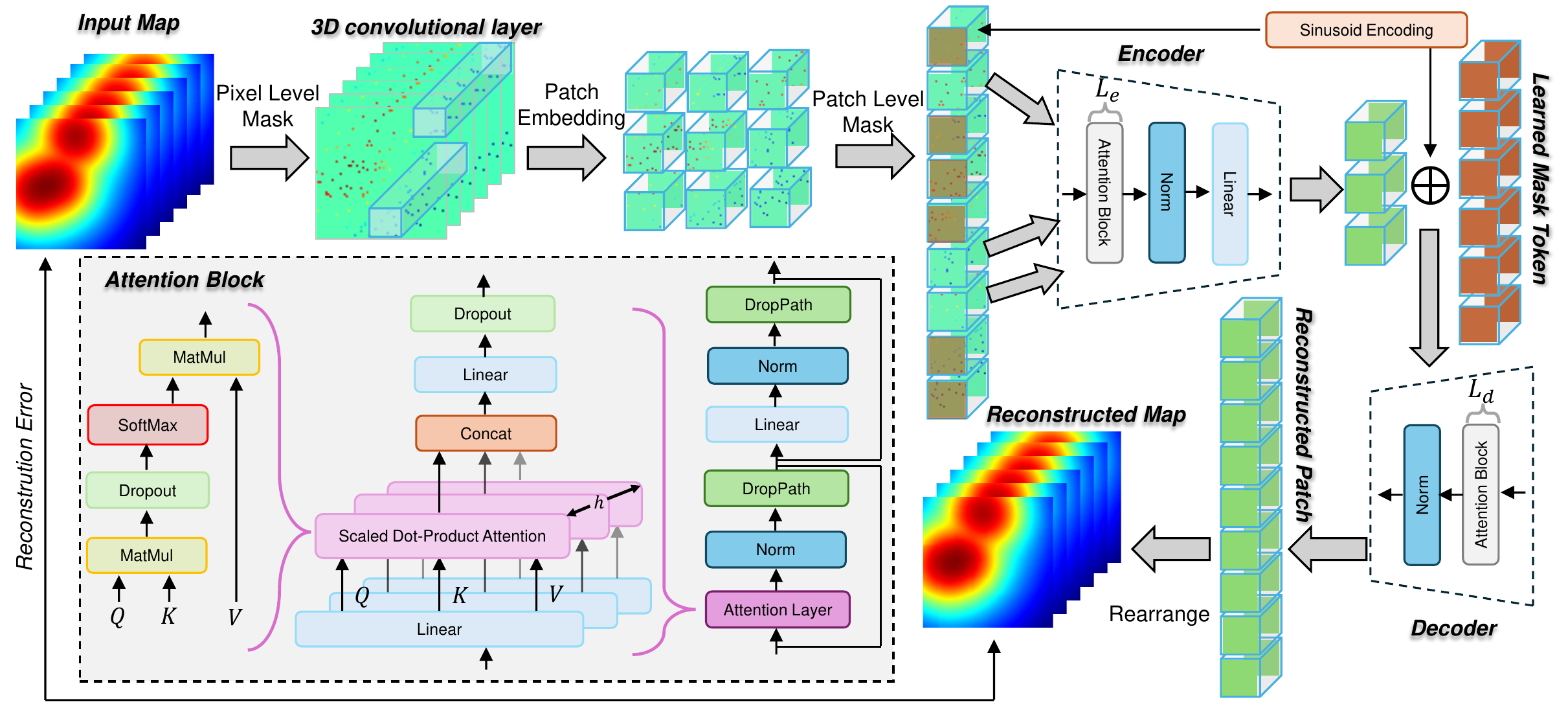}
     \caption{
 Overview of the RecMAE framework. The model processes masked spatiotemporal radio maps using 3D convolutions and patch embeddings. Visible patches are encoded with attention blocks, while masked tokens are learned and combined in the decoder to reconstruct spectrum maps. Reconstruction error guides self-supervised training.
    }
    \label{fig:recmae}
\end{figure*}

\subsubsection{Self-attention}
\label{sec:attention}

The self-attention mechanism is based on the scaled dot-product attention as introduced in Transformers \cite{vaswani2017attention}. Given a set of query vectors $\mathbf{Q} \in \mathbb{R}^{n \times d}$, key vectors $\mathbf{K} \in \mathbb{R}^{m \times d}$, and value vectors $\mathbf{V} \in \mathbb{R}^{m \times d_v}$, the attention output is a weighted sum of values, where weights are computed by the dot product of queries and keys, scaled by the dimensionality. Specifically, for a query $\mathbf{Q}$ and keys $\mathbf{K}$, the attention is: 
\begin{equation}
\mathrm{Att}(\mathbf{Q}, \mathbf{K}, \mathbf{V}) = \operatorname{softmax}\left(\frac{\mathbf{Q}\mathbf{K}^\top}{\sqrt{d}}\right) \cdot\mathbf{V},
\label{eq:attn}
\end{equation}
where $d$ is the dimensionality of the query/key vectors, and softmax denotes the softmax activation function. This operation produces an output of size $n \times d_v$ as each query attends to all key-value pairs. As shown in Fig. \ref{fig:recmae}, self-attention allows each patch token to adaptively aggregate information from other tokens, thereby capturing spatial and temporal dependencies. 


\subsubsection{Multi-Head Attention} 

To increase the expressiveness of the attention mechanism, RecMAE employs multi-head attention \cite{voita2019analyzing}. Instead of using a single attention function, the multi-head attention uses $h$ parallel attention heads. For head $i$, the queries, keys, and values are linearly projected into a subspace using learned projection matrices $\mathbf{W}_i^Q$, $\mathbf{W}_i^K$, and $\mathbf{W}_i^V$. The head output is computed as $\text{head}_i = \mathrm{Att}(\mathbf{Q} \mathbf{W}_i^Q, \mathbf{K} \mathbf{W}_i^K, \mathbf{V} \mathbf{W}_i^V)$. The outputs of all heads are then concatenated and projected again to form the final output: 
\begin{equation}
\mathrm{MHA}(\mathbf{Q}, \mathbf{K}, \mathbf{V}) = \operatorname{Concat}(\text{head}_1, \text{head}_2, \ldots, \text{head}_h)\cdot\mathbf{W}^O,
\label{eq:mha}
\end{equation}
where Concat denotes the concatenation operation, and $\mathbf{W}^O$ is the output projection matrix. Multi-head attention allows the model to attend to different representation subspaces simultaneously, which is crucial for capturing complex spatio-temporal patterns in the radio map data. In the context of RecMAE, we use multi-head self-attention within each Transformer encoder/decoder layer. 
As shown in Fig. \ref{fig:recmae}, multi-head self-attention employs multiple parallel scaled dot-product attention heads, where the original queries, keys, and values are split across heads to capture different aspects of spatial or temporal information.


\subsubsection{Patch Embedding} 
\label{subsub: patch}

To embed the spatio-temporal input for Transformer-based processing, we adopt a tubelet embedding strategy implemented via a 3D convolutional layer \cite{tong2022videomae}, as illustrated in Fig.~\ref{fig:recmae}. Let the input be a sequence of $T$ radio maps $\{\mathbf{X}_t\}_{t=1}^T$, where each frame $\mathbf{X}_t \in \mathbb{R}^{H \times W}$ represents a 2D spatial field. We first stack these frames into a 3D tensor $\mathbf{X} \in \mathbb{R}^{C \times T \times H \times W}$, where $C$ is the number of input channels.


We then divide the 3D volume into non-overlapping spatio-temporal patches of size $P_t \times P_h \times P_w$, where $P_t$ denotes the temporal patch size, i.e., the tubelet size, and $P_h$, $P_w$ are the height and width of each spatial patch. By letting $T$, $H$, and $W$ be divisible by $P_t$, $P_h$, and $P_w$, respectively, this process yields a total number of patches:
\begin{equation}
N_{patch} = \frac{T}{P_t} \cdot \frac{H}{P_h} \cdot \frac{W}{P_w}~.
\end{equation}

Each patch is projected into a $D$-dimensional latent token using a learnable 3D convolution with kernel size $(P_t, P_h, P_w)$ and stride $(P_t, P_h, P_w)$:
\begin{equation}
\mathbf{Z} = \operatorname{Conv3D}(\mathbf{X}) \in \mathbb{R}^{D \times T' \times H' \times W'}~,
\end{equation}
where $T' = T/P_t$, $H' = H/P_h$, $W' = W/P_w$, and Conv3D represents the learnable 3D convolution.
The resulting tensor is reshaped into a sequence of tokens $\mathbf{z}_n \in \mathbb{R}^{D}$ for $n = 1, \dots, N_{patch}$, which are then fed into the Transformer encoder. 

By dividing the input along both spatial and temporal dimensions, this patch embedding approach effectively captures spatial features and temporal patterns within each token, resulting in a compact and expressive representation for downstream processing.

\subsubsection{Positional Embedding} 

Since the Transformer architecture is inherently permutation-invariant with respect to the input token order, it is essential to incorporate positional information that reflects spatial structure. We employ fixed sinusoidal positional encodings \cite{vaswani2017attention}.
Let $D$ denote the embedding dimension. For a token $\mathbf{z}_n$ located at spatial position $n$ of the tokens sequence, we compute its spatial positional encoding $\mathbf{e}^{\text{spa}}_{n} \in \mathbb{R}^D$ using the sinusoidal formulation:
\begin{equation}
\begin{split}
\text{PE}_{(n, 2k)} &= \sin\left(\frac{n}{P_{PE}^{2k / D}}\right), \\
\text{PE}_{(n, 2k+1)} &= \cos\left(\frac{n}{P_{PE}^{2k / D}}\right),
\end{split}
\label{eq:sinusoid}
\end{equation}
where $n$ is the position index, $k = 0, 1, \ldots, D/2 - 1$, $P_{PE}$ is a predefined number usually set as 10000, and $\mathbf{e}^{\text{spa}}_{n} = [PE_{(n,0)}, \ldots, PE_{(n,D-1)}]$. 

Each final token input to the Transformer encoder is the sum of the patch embedding and its positional encodings:
\begin{equation}
\mathbf{y}_{n} = \mathbf{z}_{n} + \mathbf{e}^{\text{spa}}_{n}~. 
\label{eq:pos_enc}
\end{equation}
The spatio-temporal patch embedding and positional encodings help the model reason over both spatial structures and temporal dynamics. The same positional encodings are applied in both the encoder and decoder to preserve alignment across the two stages.



\subsubsection{Model Architecture} 

\textbf{Masking and Encoder:} 
Following the MAE paradigm~\cite{he2022masked}, a subset of the tokens is randomly masked out and omitted from the encoder input. 
We define a binary mask over the set of all tokens. 
Let $\mathcal{M}$ denote the set of indices of masked patches and $\mathcal{V}$ denote the set of visible (unmasked) patch indices, such that $\mathcal{M} \cup \mathcal{V}$ includes all patches, i.e., $|\mathcal{M}| + |\mathcal{V}| = N_{patch}$. Typically, a high masking ratio $r_{patch}$ is used, e.g., mask $70\%-90\%$ of tokens \cite{he2022masked}, so that the model learns to infer a large portion of missing data from a small portion of visible context. 
As shown in Fig. \ref{fig:recmae}, the encoder receives as input only the tokens corresponding to $\mathcal{V}$. 
Namely, we feed $\{\mathbf{y}_{n_v}, n_v \in \mathcal{V}\}$ into a Transformer encoder consisting of $L_e$ layers of multi-head self-attention and feed-forward network blocks. Through these layers, each visible token's representation is updated by attending to other visible tokens. Let $\mathbf{h}_{n_v}$ denote the encoded representation of a visible token $\mathbf{y}_{n_v}$ after the final encoder layer. 
Note that tokens with indices in $\mathcal{M}$ are invisible tokens and are the input to the encoder in the proposed framework.


\textbf{Decoder and Reconstruction:} 
The decoder network aims to reconstruct the original radio map from the encoded visible tokens and the masked tokens. 
To this end, we introduce the masked patches into the token sequence by adding a learned mask token for each index in $\mathcal{M}$. 
Specifically, for each masked patch index $n_m \in \mathcal{M}$, we create a token \[\mathbf{h}_{n_m} = \mathbf{m} + \mathbf{e}^{\text{spa}}_{n_m},\] where $\mathbf{m} \in \mathbb{R}^{D}$ is a learned mask vector shared for all masked positions, and we add the same positional encodings as used in the encoder to indicate where this token belongs. 
For each visible patch $n_v \in \mathcal{V}$, we take its encoded representation $\mathbf{h}_{n_v}$ from the encoder and also add the positional encodings $\mathbf{e}^{\text{spa}}_{n_v}$. 
This combined set of tokens, of size $N_{patch}$ covering both originally visible and masked patches, is then passed through the Transformer decoder, which consists of $L_d$ layers of multi-head self-attention and feed-forward blocks. 
In the decoder, tokens can attend to both original visible tokens and the mask tokens, allowing information to flow from observed regions to infer missing regions. 

The output of the decoder is a set of $N_{patch}$ decoded vectors, i.e., $\{\mathbf{o}_{n}\}$, one for each patch position at each time. 
The final step is to map these output vectors to reconstructed patch values. We apply a linear projection that inverts the patch embedding, producing $\hat{\mathbf{z}}_{n} \in \mathbb{R}^{P_t\times P_h \times  P_w}$ from $\mathbf{o}_{n}$. 
In other words, the decoder outputs are transformed to the same dimensionality as the flattened patch input, yielding a reconstructed patch $\mathbf{\hat{Z}}_n$ for both originally visible and masked patches. The set of all reconstructed patches $\{\mathbf{\hat{X}}_{n}\}_{n=1}^{N_{patch}}$ can be reassembled into $T$ full reconstructed frames, which we denote as $\mathbf{\hat{X}}_1, \mathbf{\hat{X}}_2, \ldots, \mathbf{\hat{X}}_T$, as present in Fig. \ref{fig:recmae}. 

\subsubsection{Dual-mask Mechanism} 

To robustly reconstruct radio maps from limited sensor measurements, we propose a dual-masking strategy that conceals information at two distinct scales: the patch (token) level and the pixel level. 
The pixel-level masking simulates limited sensor coverage by randomly dropping out individual pixel measurements in the input radio map, reflecting real-world scenarios where sensor readings are available only at certain spatial locations
On top of this, the patch-level mask (as discussed in the above RecMAE training) is applied to withhold entire regions from the encoder \cite{he2022masked}. 
By combining fine-grained pixel masking with coarse patch masking, the model is forced to learn both local and global context to fill in missing signal values.
Consequently, the RecMAE decoder aims to reconstruct the original unmasked radio map, simulating the task of accurately completing data from limited sensor measurements.
This dual-masked mechanism thus learns to infer a complete radio map from highly incomplete data, improving its reconstruction ability under sensing limitations.

Formally, considering a sequence of $T$ radio maps $\{\mathbf{X}_t\}_{t=1}^T$, we define a binary pixel mask tubelet $\textbf{M}^{pixel} \in \{0,1\}^{T\times H\times W}$, where $\textbf{M}^{pixel}(t, i,j)=1$ indicates that the pixel at location $(i,j)$ is observed (available) at time $t$, $\textbf{M}^{pixel}(t, i,j)=0$ means the pixel’s value is missing to simulate lack of sensor data. Moreover, the pixel mask radio is $r_{pixel}$, which is usually chosen based on the number of sensors.
Since the sensor positions are fixed within a certain time slot, the mask values are held consistent along the temporal dimension, i.e., 
\begin{equation}
    \textbf{M}^{pixel}(t, i,j) = \textbf{M}^{pixel}(t', i,j),~ \forall t,t'\in[1,T].
\end{equation}
The pixel-masked input $\textbf{X}^{pixel}$ is obtained by element-wise applying this mask to the original image:
\begin{equation}
\label{eq:pixelmask}
    \textbf{X}^{pixel} = \textbf{M}^{pixel}\circ\textbf{X}.
\end{equation}
Next, the pixel-masked data $\textbf{X}^{pixel}$
  is divided into patches, and subsequently, a patch-level mask is applied to these patches before feeding them into the encoder, as we introduced above.

\subsubsection{Loss Functions} 

Training of RecMAE uses a loss function that penalizes reconstruction errors in both space and time. Therefore, 
the training objective is to make the stacked reconstructed frames $\mathbf{\hat{X}}$ as close as possible to the ground-truth $\mathbf{X}$ despite both pixel and patch level masks.
In practice, we compute the reconstruction loss using the mean squared error (MSE) between reconstructed patches and their corresponding original patches, defined as follows:
\begin{equation} 
Loss = \frac{1}{N_{patch}} \sum_{i=1}^{N_{patch}} 
\left| \hat{\mathbf{z}}_i - \mathbf{z}_i \right|^2~, 
\label{eq:loss_spatial} 
\end{equation}
which directly measures the reconstruction error since the patch-to-image transformation involves only reshaping operations without additional data transformations.
By minimizing $Loss$, the RecMAE model learns to accurately fill in missing spatial data in each frame while also ensuring that the reconstructed sequence of frames is temporally coherent. This loss formulation drives the encoder-decoder to learn meaningful spatio-temporal representations of the radio map data, enabling effective reconstruction even under high masking ratios.

\subsubsection{Reconstruction Inference}

The reconstruction process in the RecMAE framework involves encoding observed spatio-temporal data into latent representations and subsequently decoding these representations to reconstruct the complete spatio-temporal sequence. 
Given that sensor observations inherently provide partial information, 
the input data $\mathbf{\tilde{P}}^i$ already incorporates a pixel-level mask $\mathbf{W}^{i}$. Subsequently, the collected data undergo an additional patch-level masking step to form the encoder's input. 
In the decoding stage, the latent representations are utilized to reconstruct the entire radio map.
Leveraging the GenAI paradigm of unsupervised learning, transformer-based decoder blocks iteratively refine the reconstruction, progressively recovering complete frames from partially masked inputs.

The complete GenAI reconstructor algorithm is detailed in Algorithm \ref{alg:recmae}.


\begin{algorithm}[tpb]
    \small 
    \caption{\textcolor{black}{RecMAE Training and Inference}}
    \label{alg:recmae}
    $\mathbf{Input}$: Encoder network $f_{\theta}$, Decoder network $g_{\phi}$, Radio Map Dataset $\mathcal{D}$, Masking ratio $r_{patch}$ and $r_{pixel}$, Learning rate $\gamma$, \# of training epochs $N_e$;\\
    \vspace{0.5em}
    \textit{Procedure 1: RecMAE Training};\\
    \quad \For{epoch $= 1, 2, \dots, N_e$}{    
    
    \quad Sample a batch of stacked radio data $\mathbf{X}$ from dataset $\mathcal{D}$\\
    \quad Generate masked data $\mathbf{X}^{pixel}$ by Eq. \eqref{eq:pixelmask} with ratio $r_{pixel}$\\
    \quad Encode visible patches by encoder: $\mathbf{Z} = f_{\theta}(\mathbf{X}^{pixel})$\\
    \quad Generate masked patch indices $\mathcal{V}$ by randomly masking patches with ratio $r_{patch}$\\
    \quad Decode latent representation by decoder: $\hat{\mathbf{X}} = g_{\phi}(\mathbf{Z})$\\
    \quad Compute reconstruction loss in Eq. \eqref{eq:loss_spatial}\\
    \quad Update parameters $\theta, \phi$ by gradient descent using learning rate $\gamma$\\
    }
    \vspace{0.5em}
    \textit{Procedure 2: RecMAE Inference};\\
    \quad Given an observed spatio-temporal radio data $\mathbf{X}$\\
    \quad Generate masked patch indices $\mathcal{V}$ by randomly masking patches with ratio $r_{patch}$\\
    \quad Encode visible patches: $\mathbf{Z} = f_{\theta}(\mathbf{X})$\\
    \quad Decode latent representation: $\hat{\mathbf{X}} = g_{\phi}(\mathbf{Z})$\\
    $\mathbf{Output}$: Reconstructed image $\hat{\mathbf{X}}$
\end{algorithm}


\subsection{Multi-agent Spectrum Cartography}




In our multi-agent spectrum cartography scenario involving dynamic UAV sensors, each UAV can only collect radio signal information within a limited sensing range from its current vantage point. Consequently, the true environmental state, the spatial distribution of RF power, which is typically critical for determining the UAV's next position, cannot be fully observed by any single agent at any given time slot.
This discrepancy between the global state and each UAV’s local observations results in partial observability: \textit{they never have complete knowledge of the underlying state.}
Instead, each UAV must make decisions under uncertainty, using only incomplete observations of the radio map. This is the reason for the complexity of the temporal spectrum,
as the UAVs must infer unobserved spectrum conditions from limited RSSI data and coordinate their exploration of the environment over time.

\subsubsection{Partially Observable Markov Decision Process}

We model the multi-UAV trajectory planning task as a partially observable Markov decision process (POMDP) \cite{spaan2012partially} to capture the sequential decision-making under incomplete observations. A POMDP is defined by the tuple $(\mathcal{S}, \mathcal{A}, \mathcal{O}, P, O, R, \gamma)$, where:
\begin{itemize}
    \item $\mathcal{S}$ (State Space): The set of all possible environment states. 
    A state $s\in \mathcal{S}$ represents the true radio spectrum map at the current time.

    \item $\mathcal{A}$ (Action Space): The set of actions that the agent can take. For a single UAV, an action $a\in \mathcal{A}$ controls the UAV’s next position.
    
    
    \item $\mathcal{O}$ (Observation Space): The set of possible observations. An observation $o\in \mathcal{O}$ corresponds to the local measurement RSSI data the UAV receives from its sensors during a time slot.  
    
    
    \item $P(s' \mid s, a)$ (Transition Probability): The state transition model, defining the probability of moving to a new state $s'$ when the agent takes action $a$ in state $s$. 
    
    
    \item $O(o\mid s', a)$ (Observation Function): The observation model, giving the probability of receiving observation $o$ after taking action $a$ and ending up in state $s'$. 
    
    
    \item $R(s, a)$ (Reward Function): The immediate reward obtained by the agent for taking action $a$ in state $s$. We design the reward to encourage accurate and efficient mapping of the spectrum based on reconstruction error. 
    
    \item $\gamma$ (Discount Factor): A factor $0 \le \gamma < 1$ that discounts future rewards relative to immediate rewards.
    
\end{itemize}


Under this POMDP formulation, the UAV agent operates in time slots, i.e., $T_1,\dots, T_{n_t}$. At each time slot $t$, the environment is in some hidden state $s_t \in \mathcal{S}$. The UAV does not directly observe environment state $s_t$; instead, it receives an observation $o_t \in \mathcal{O}$ correlated with $s_t$.
Based on this local observation $o_t$, the UAV chooses an action $a_t \in \mathcal{A}$ to move to a new position. 


\subsubsection{Multi-agent POMDP}

For multiple UAVs, we consider $M_d$ dynamic UAV agents, operating simultaneously, which extends the POMDP to a multi-agent POMDP setting. 
Each UAV $i$ receives its own observation $o^i_t \in \mathcal{O}_i$, where $\mathcal{O}_i$ is the observation space for agent $i$. In practice, $o^i_t$ consists of the sensor readings UAV $i$ collects at time $t$.
Due to the limited range, $o^i_t$ only depends on a local slice of the state $s_t$, and may differ from $o^j_t$ of another UAV $j$.
No single agent can observe the state of full positions, but collectively their observations ${o^1_t,\dots,o^N_t}$ provide complete information about the spectrum. 
Each UAV makes its decision decentrally based on its own observation. Agent $i$ chooses an action $a^i_t \in \mathcal{A}_i$, from its action space $\mathcal{A}_i$. 
Then, all agents share the same reward function $R$, defined on the global state $s_t$ and joint action $\mathbf{a}_t = (a^1_t, a^2_t, \dots, a^N_t)$. This shared reward encourages the UAVs to collaborate to improve the overall spectrum map.



\subsubsection{Multi-agent Reinforcement Learning}

We formulate this cooperative multi-UAV trajectory optimization problem as a multi-agent reinforcement learning (MARL) task under partial observability. 
Each UAV $i$ employs a policy $\pi^i$ that maps its observation history to actions. 
During execution, the policy is conditioned on the current observation $o^i_t$ for decision-making. 
The policies are decentralized in that there is no single controller; each agent makes its own decision independently based on local information. 
However, the learning of these policies is centralized or coordinated during training to encourage cooperation. 

The objective for the team of agents is to find a set of policies $(\pi^1,\pi^2,\dots,\pi^N)$ that maximizes the expected cumulative reward, equivalently, minimizes the long-term mapping error. 
In formal terms, by letting $\mathbf{\pi} = (\pi^1,\ldots,\pi^N)$ denote the collection of policies, the MARL optimization can be written as follows:
\begin{equation}
\pi^* = \arg\max_{\pi^1, \dots, \pi^N} \mathbb{E}\left[ \sum_{t=0}^{\infty} \gamma^t R\left(s_t, a_t^1, \dots, a_t^N\right) \right],
\label{eq:policy}
\end{equation}
subject to each agent $i$ choosing actions according to its policy, $a^i_t \sim \pi^i(o^i_t)$, at every time step.

\subsection{Multi-agent Diffusion Policy}


\begin{figure}[htbp]
    \centering
    \includegraphics[width= 0.98\linewidth]{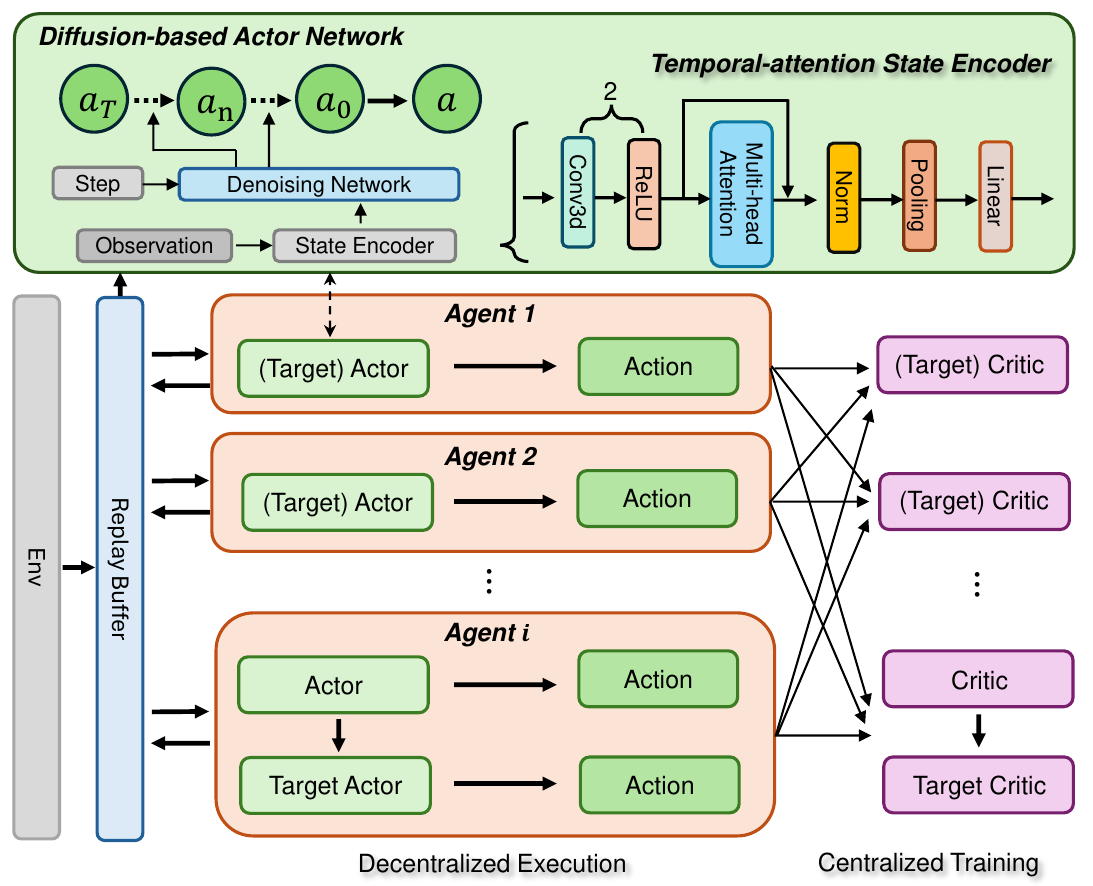}
    \caption{
    Overview of the MADP framework. The framework combines a diffusion-based actor network with a temporal-attention state encoder to guide multi-agent policy learning. Each agent selects actions independently via decentralized execution, while a centralized critic network enables training using shared information. 
    }
    \label{fig:actor}
\end{figure}

For the GenAI planner, we propose an MADP that extends the generative diffusion model optimization framework \cite{du2024enhancing} in multi-agent scenarios.

\subsubsection{Diffusion Process Background}

We build upon denoising diffusion probabilistic models (DDPMs) to enrich the policy representation. A DDPM includes a forward diffusion process that gradually adds noise to data, and a learned reverse process that removes noise to recover data samples \cite{ho2020denoising}.

Formally, let $x_0$ represent an input action data, which is considered to be a sample from the optimal action distribution, and $a_n$ its noisy version after $n$ diffusion steps. 
The forward process is a Markov chain 
\begin{equation}
q(x_1, \dots, x_T \mid x_0) = \prod_{n=1}^{T} q(x_n \mid x_{n-1})
\end{equation} 
with Gaussian transitions 
\begin{equation} 
q(x_n \mid x_{n-1}) = \mathcal{N}\Big(x_n ; \sqrt{\alpha_n}x_{n-1},~ (1-\alpha_n) \mathbf{I}\Big), 
\label{eq:forward_diff} 
\end{equation} 
for $n=1,2,\dots,T$, where $\alpha_n \in (0,1)$ is a variance schedule. After the $T$ steps, $a_T$ is approximately distributed as a normal Gaussian distribution. 

The generative reverse process is a parametric Markov chain 
\begin{equation}
p_\theta(x_{0:T}\mid c) = p(x_T)\prod_{n=1}^T p_\theta(x_{n-1}\mid x_n,c),
\end{equation}
which learns to invert the diffusion. Here $c$ denotes conditioning information, which is the observation $o\in\mathcal{O}$ in our case.
Each reverse step is modeled by a Gaussian 
\begin{equation} 
p_\theta(x_{n-1}\mid x_n,c) = \mathcal{N}\big(x_{n-1}; \mu_\theta(x_n, c, n), \Sigma_\theta(x_n, c, n)\big), 
\label{eq:reverse_diff} 
\end{equation} 
where $\mu_\theta$ and $\Sigma_\theta$ are the predicted mean and covariance at step $n$ given input $a_n$ and condition $c$. 



The denoising network $\epsilon_\theta$ is trained by minimizing a weighted sum of mean-squared errors at each diffusion step. 
The objective is: 
\begin{equation} 
Loss_{\text{diffusion}}(\theta) = \mathbb{E}_{n,x_0,\epsilon}\Big[ \big| \epsilon - \epsilon_\theta(\sqrt{\bar{\alpha}_n}x_0 + \sqrt{1-\bar{\alpha}_n }\epsilon,c,n ) \big|^2 \Big], 
\label{eq:diffusion_loss} 
\end{equation} 
where $\bar{\alpha}_n = \Pi_{i=1}^n\alpha_n$.
The loss function encourages $\epsilon_\theta(x_n,c,n)$ to correctly predict the noise $\epsilon$ added to the clean input $x_0$ at every step. 
By learning this reverse diffusion model, the network can generate sample actions $x_0$ from pure noise $x_T\sim \mathcal{N}(0,\mathbf{I})$ via iterative denoising. 


\subsubsection{Diffusion-Based Actor Network}
\label{sec:tattn}

In our MADP architecture, each agent’s actor policy is implemented as a conditional denoising diffusion model framework.
This means that instead of outputting an action in one forward pass, the actor generates actions by simulating the denoising conditioned on the agent’s state.

Specifically, let $o^i_t$ be the observation for agent $i$ at time $t$. The actor draws an initial noise $x_{T}^i \sim \mathcal{N}(0,\mathbf{I})$ and iteratively applies the learned denoising mapping $T$ times: 
\begin{equation}
x_{n-1}^i = \frac{1}{\sqrt{\alpha_t}}\left(x^i_t - \frac{1 - \alpha_t}{\sqrt{1 - \bar{\alpha}_n}}\boldsymbol{\epsilon}_\theta(x^i_n,o_{t}^i, n)\right) + \sigma_n z_n,
\label{eq:denoise_update} 
\end{equation}
where $n=T, T-1, \dots, 1$, $\sigma_n = \sqrt{1-\alpha_n}$,$z_n\sim \mathcal{N}(0,\mathbf{I})$ is added noise, and $\epsilon_\theta^i$ is the actor network for agent $i$.

After $T$ reverse steps, we obtain $x_{0}^i$, which is the action $a^i_t$ for agent $i$ at time $t$. 
In other words, $\pi^i_{\theta}(o^i_{t}) \equiv a^{i}_{t}$ is generated via $a^i_t = x_{0}^{i}$ with $x_{0}^{i}$ produced by the diffusion model $\epsilon_\theta^i$ conditioned on $o_{t}^i$. 
This procedure can be interpreted as the agent refining a random initial noise, into a coherent action using the learned denoising policy. 
The conditioning $o_{t}^i$, which contains the agent’s local observation, guides the denoising network at each step, allowing the actor to generate state-dependent actions. 
 Thus, the diffusion-based actor update encourages actions that both yield high Q-values and lie on the manifold of the learned action distribution.



\textbf{Temporal-attention State Encoder:} 
To effectively condition the diffusion policy on relevant history, we introduce a temporal-attention state encoder for each agent, as illustrated in Fig. \ref{fig:actor}. 
This encoder, denoted $g_\psi$, produces an enhanced state representation $h_{t}^i$ that captures temporal context from a sequence of observations via the attention mechanism introduced in Section \ref{sec:attention}. 
Specifically, $g_\psi$ is implemented using a CNN augmented with a temporal self-attention mechanism to effectively capture sequential dependencies and emphasize informative time steps.
This design helps the encoder approximate the underlying state by incorporating temporal information, which is especially useful under partial observability in POMDP. 




\subsubsection{MADP Framework}

We follow the multi-agent deep deterministic policy gradient framework (MADDPG) to train our MADP. Consider an $M_d$-agent POMDP defined by the tuple $(\mathcal{S},\{\mathcal{A}\}_{i=1}^{M_d}, \{\mathcal{O}\}_{i=1}^{M_d}, P, O, R, \gamma)$.
Each agent $i$ has a diffusion-based actor $\pi^i_{\theta}$ and a Q-function critic $Q^i_{\psi}$. 
Under decentralized execution, agent $i$ chooses actions $a^i_{t} = \pi^i_{\theta}(o^i_{t})$ based only on its own observation $o^i_{t}$. 

During centralized training, the critic for agent $i$ is augmented with global information, which takes as the collection of all agents’ observations $\mathbf{o}_t = (o^i_{t},\dots,o^{M_d}_{t})$ and the joint action $\mathbf{a}_t=(a^i_{t},\dots,a^i_{t})$. 
The critic $Q^i_{\psi}(\mathbf{o}_t, \mathbf{a}_t)$ estimates cumulative discounted reward for agent $i$ starting from state $s_t$ after all agents execute actions $\mathbf{a}_t$ and follow policy $\Pi^i_\theta$ thereafter. 
Each agent aims at maximizing a cooperative team reward over $n_T$ time slots, defined as follows:
\begin{equation}
    R_i = R = \mathbb{E}\bigg[\sum_{t=0}^{n_T-1} \gamma^t r(s_t, \mathbf{a}_t)\bigg],
\end{equation}
where $r(s_t, \mathbf{a}_t)$ represents the reconstruction error as computed by the GenAI reconstructor in time slot $t$.


The MADP training alternates between updates of the critics ${\psi}$ and actors ${\theta}$ using experiences from a replay buffer, as present in Fig. \ref{fig:actor}. 
For a given transition $s = (s_t, \mathbf{a}_t, r(s_t, \mathbf{a}_t), s_{t+1})$ sampled from the replay buffer $\mathcal{D}$, the critic for agent $i$ is updated by minimizing the reward error: 
\begin{equation} 
Loss^i_{\text{critic}}(\phi_i) = \mathbb{E}\Big[ \big(Q^i_{\psi}(\mathbf{o}_t, \mathbf{a}_{t}) - y^i_{t}\big)^2 \Big],
\label{eq:critic_loss} 
\end{equation} 
and 
\begin{equation}
    y^i_{t} = r(s_t, \mathbf{a}_t) + \gamma Q'^i_{\psi}(\mathbf{o}_{t+1}, \mathbf{a}'^i_{t+1}), 
\end{equation}
where 
$\mathbf{a}'^i_{t+1} = \pi^i_{\theta}(o_{j,t+1})$ is the next action for agent $i$ given by the target actor network, which is a delayed copy of $\pi_{\theta}^j$, and $Q'$ is the target critic network \cite{lowe2017multi}.
To maximize the critic’s estimate of the return, we then define the actor loss for each agent $i$ as \cite{silver2014deterministic}:
\begin{equation} 
Loss^i_{\text{actor}}(\theta_i) = -\mathbb{E}_{s\sim\mathcal{D}}\Big[ Q^i_{\phi}\big(\mathbf{o}, \mathbf{a}) \Big], 
\label{eq:actor_loss} 
\end{equation} 
where $\mathcal{D}$ is the replay buffer distribution.

\subsubsection{MADP Execution Process}

During the inference phase, the learned policies of each agent are deployed in a decentralized manner. Specifically, at each time step $t$, each agent $i$ receives its local observation $o_i^t$. Then, the agent computes its action with diffusion-based actor network
$ a_i^t = \pi^i_\theta(o_i^t)$. All agents simultaneously execute their actions in the environment, resulting in a new joint state and individual observations for the next time step. This process repeats until the episode terminates.
Unlike the training phase, which leverages centralized training with access to the collection of observations and actions of all agents, the inference process operates in a fully decentralized execution setting, relying solely on individual observations. 
This enables agents to act independently in real-time scenarios while benefiting from the coordinated policies learned during training.
From the computational complexity, the primary overhead of the proposed MADP model lies in its multi-step diffusion process. As each individual step incurs a similar computational cost to that of traditional methods, the overall complexity is approximately $T$ times, where $T$ denotes the number of denoising steps.

In summary, the complete GenAI planner algorithm is detailed in Algorithm \ref{alg:madp}.

\begin{algorithm}[tpb]
    \small
    \caption{\textcolor{black}{MADP Training and Execution}}
    \label{alg:madp}
    $\mathbf{Input}$: Actor networks $\{\mu_{\theta_i}\}_{i=1}^{N}$, Critic networks $\{Q_{\phi_i}\}_{i=1}^{N}$, Replay buffer $\mathcal{D}$, Learning rate $\gamma$, Discount factor $\delta$, Soft update parameter $\tau$, Number of agents $M_d$, \# of episodes $N_e$, \# of time slots $n_T$;\\
    \vspace{0.5em}
    \textit{Procedure 1: MADP Training};\\
    \quad \For{episode $= 1, 2, \dots, N_e$}{
    \quad Initialize environment $\{s_i\}_{i=1}^{n_t}$ and observe initial states $\{o_0^i\}_{i=1}^{N}$\\
    \quad \For{$t = 0, 1, \dots, n_T-1$}{
    \quad \quad Each agent $i$ selects action $a^i_t = \mu_{\theta}^i(o^i_t)$\\
    \quad \quad Execute joint action $\mathbf{a}_t = (a^1_t, \dots, a^N_t)$ and observe $r(s_t,\mathbf{a}_t)$\\
    \quad \quad Store transition $(s_t, \mathbf{a}_t, r(s_t, \mathbf{a}_t), s_{t+1})$ in buffer $\mathcal{D}$\\
    \quad \quad Sample minibatch from buffer $\mathcal{D}$\\
    \quad \quad \For{each agent $i=1, \dots, N$}{
    \quad \quad \quad Compute target action $\mathbf{a}'_i = \pi'^i_\theta(o^i_t)$\\
    \quad \quad \quad Compute target Q-value: $y^i_{t} = r(s_t, \mathbf{a}_t) + \gamma Q'^i_{\psi}(\mathbf{o}_{t+1}, \mathbf{a}'_{t+1})$\\
    \quad \quad \quad Update critic by minimizing loss: $L = \left(Q^i_{\phi}(\mathbf{s}, \mathbf{a}) - y_i\right)^2$\\
    \quad \quad \quad Update actor by minimizing loss: $Loss^i_{\text{actor}}(\theta_i) = -\mathbb{E}_{s\sim\mathcal{D}}\Big[ Q^i_{\phi}\big(\mathbf{o}, \mathbf{a}) \Big], $\\
    \quad \quad \quad Soft update target networks: $\theta_i' \leftarrow \tau \theta_i + (1 - \tau) \theta_i'$ and $\phi_i' \leftarrow \tau \phi_i + (1 - \tau) \phi_i'$\\
    \quad \quad }
    \quad }
    }
    \vspace{0.5em}
    \textit{Procedure 2: MADP Execution};\\
    \quad Given current environment states $s_t$\\
    \quad Each agent $i$ selects action $a^i_t = \phi^i_{\theta}(o^i_t)$\\
    $\mathbf{Output}$: Joint action $\mathbf{a}_t = (a^1_t, \dots, a_t^N)$
\end{algorithm}












\section{Numerical Results}\label{sec:sec5}

In this section, we evaluate the proposed two-stage GenAI spectrum cartography framework through extensive experiments. Specifically, we assess the performance of both components: the GenAI Reconstructor (RecMAE) and the GenAI Planner (MADP). We also provide a detailed analysis of the experimental results.
The experiments are conducted on a server equipped with three NVIDIA RTX A6000 GPUs, an Intel(R) Xeon(R) Silver
4410Y 12-core processor. The system runs Ubuntu 22.04
operating system and utilizes PyTorch for implementation.

\subsection{Performance Analysis of the GenAI Reconstructor Stage}


To evaluate the performance of the proposed RecMAE, we consider a temporal spectrum reconstruction scenario with $n_T = 10$ discrete time slots. In each slot, sensors are randomly deployed based on a coverage level characterized by the sensing ratio $\rho$. The reconstruction quality is assessed using the mean squared error (MSE) between the predicted and ground truth radio power maps.

\textbf{Baseline Methods:}
We compare the proposed RecMAE against several baseline methods, including classical interpolation and deep learning models. 
The key methods and their abbreviations are as follows:
\begin{itemize}
    \item[-] \textbf{AE}: Autoencoder baseline for spectrum completion \cite{teganya2021deep}. Note that this baseline was not originally designed for temporal data, and we made certain modifications to adapt it accordingly.
    \item[-] \textbf{Kriging}: A kernel-based Kriging interpolation method adopted as a non-learning baseline for spatial spectrum completion \cite{boccolini2012wireless}.
\end{itemize}

\textbf{Experiment Settings:}
In this study, we simulate a low-altitude economic activity scenario using the key parameters listed in Table \ref{tab:experimental_settings}. The simulated area is a dense urban environment, and all propagation characteristics are taken directly from the 3GPP TR 38.901 v16.1.0 urban model \cite{3GPP38901}.
We set the sensing UAV’s altitude at $h_{\text{sensor}} = 50$ m, a height that both preserves near-ground spatial resolution and secures reliable LOS links, making it well-suited for accurate spectrum-map reconstruction \cite{chen2024high}.
The speed of mobile users is set within the range of $[1.0, 1.5]\,\mathrm{m/s}$ to mimic human walking behavior.
Specifically, we consider a crossroad scenario located at the center of the region, where the simulated humans are walking along the road.
We generate 1,000 training samples and 200 test samples following Eq. \eqref{eq:generateDBM}. The learning models are trained for 2,000 epochs using a sensing ratio of $\rho = 10\%$, and evaluated under three different sensing ratios: $\rho \in \{10\%, 5\%, 3\%\}$.




\begin{table}[ht]
    \centering
    \setlength{\tabcolsep}{2mm}\footnotesize
    \caption{Experimental settings.}
    \begin{tabular}{c|c||c|c}
        \hline
        \textbf{Parameter} &  \textbf{Value} &\textbf{Parameter} &  \textbf{Value} \\
        \hline
        $h_{\text{sensor}}$ \cite{chen2024high} & $50\,\mathrm{m}$ & $n_{\text{LOS}}$, $n_{\text{NLOS}}$ \cite{3GPP38901}  & $2.2$, $3.8$  \\
        $X$  & $256\,\mathrm{m} \times 256\,\mathrm{m}$ & $P_{\text{UE}_i}^{\text{dBm}}$  & $20\,\mathrm{dBm}$  \\
        $\Delta x$, $\Delta y$  & $4\,\mathrm{m}$, $4\,\mathrm{m}$ & $N$  & $[3 ,5]$ \\
        $H$, $W$  & $64$, $64$ & $T_i$ & $16$\\
        $f $&$ 1.8\,\mathrm{GHz}$& $h$ & $12$ \\
        $L_e$, $L_d$ & $12$, $12$ & $P_t, P_h, P_w$ & 2,8,8\\
        $r_{patch}$, $r_{pixel}$ & 0.75, 0.90 &  $\sigma_{\text{NLOS}}$ \cite{3GPP38901} & $6\,\mathrm{dB}$\\
        $d_{\text{corr}}$ \cite{3GPP38901} & $50\,\mathrm{m}$ & $a_{\text{LOS}}, b_{\text{LOS}}$ \cite{alzenad20173} & 9.61, 0.16\\
        \hline
    \end{tabular}
    \label{tab:experimental_settings}
\end{table}

 \begin{table}[h]
    \centering
    \setlength{\tabcolsep}{1mm}\footnotesize
    \caption{
    MSE and inference time (in second (s)) on different sensing radio $\rho$. ``NA" indicates that no repeated experiments were conducted."}
    \begin{tabular}{c|c|c|c|c}
    \hline
        \multirow{1}{*}{Model} & ~& \multicolumn{1}{c|}{RecMAE (Ours)} & \multicolumn{1}{c|}{AE} &  \multicolumn{1}{c}{Kriging}\\
        \hline
        $\rho$ &  MSE &  \textbf{0.39 $\pm$ 0.00} & 0.56 $\pm$ 0.00 & 0.54 $\pm$ NA\\
        10\% &Time & 24.99 $\pm$ 1.56 & \textbf{5.81 $\pm$ 0.28} & 10973.5 $\pm$ NA\\\hline
        $\rho$ & MSE &\textbf{ 0.53 $\pm$ 0.01} & 1.88 $\pm$ 0.05 &  1.65 $\pm$ NA \\
        5\%& Time & 25.13 $\pm$ 0.22 & \textbf{5.76 $\pm$ 0.08} & 10191.4 $\pm$ NA \\\hline
        $\rho$ & MSE & \textbf{0.90 $\pm$ 0.02} & 7.95 $\pm$ 0.14 &  2.11 $\pm$ NA \\
        3\% &Time & 25.89 $\pm$ 1.37 & \textbf{5.80 $\pm$ 0.16} & 8754.06 $\pm$ NA\\
        \hline
    \end{tabular}
    \label{tab:mse_comparison}
\end{table}

\textbf{Evaluation:}
Table \ref{tab:mse_comparison} summarizes the average reconstruction error and runtime for each method under different sensing ratios $\rho$. The proposed RecMAE consistently achieves the lowest MSE across all coverage levels. For example, when $\rho = 10\%$, corresponding to more abundant sensors, RecMAE achieves an MSE of only 0.39, substantially outperforming Kriging with an MSE of approximately 0.54, and the AE baseline, which reaches a MSE of around 0.56 under the same conditions.
At the lowest coverage level of $\rho = 3\%$, RecMAE still maintains a low MSE of about 0.90, compared to 2.11 for Kriging and 7.95 for the AE.
Quantitatively, the proposed approach achieves approximately an 88.68\% reduction in reconstruction error compared to AE and a 57.35\% reduction compared to Kriging.
These results highlight that RecMAE delivers the highest reconstruction accuracy and lower standard deviation and remains particularly robust in highly sparse regimes where the baseline deep model fails. 
Even when trained under the same settings as the baseline at $\rho = 10\%$, RecMAE consistently maintains superior performance without requiring any modification to the testing setup, demonstrating strong robustness and generalization across varying sensing levels.

Regarding inference time, RecMAE takes approximately 24 to 26 seconds to reconstruct the full map over the entire test dataset. 
While the inference time is about four times longer than the AE baseline, which takes around 5.5 to 6.0 seconds, it is quite reasonable considering the RecMAE’s accuracy gains. 
Moreover, when averaged across individual time slots, the time per frame is comparable to that of the AE model,
In contrast, Kriging is significantly slower, as it computes the covariance matrix separately for each frame \cite{boccolini2012wireless}. As a result, it requires approximately $1.0 \times 10^4 $ seconds or several hours to interpolate the entire test dataset, making it computationally prohibitive for large-scale or repeated experiments. This substantial time cost is also the reason why we did not perform multiple trials for Kriging (i.e., ``NA'' in Table \ref{tab:mse_comparison}).

\begin{figure*}[htbp]
    \centering
    \includegraphics[width= 0.80\linewidth]{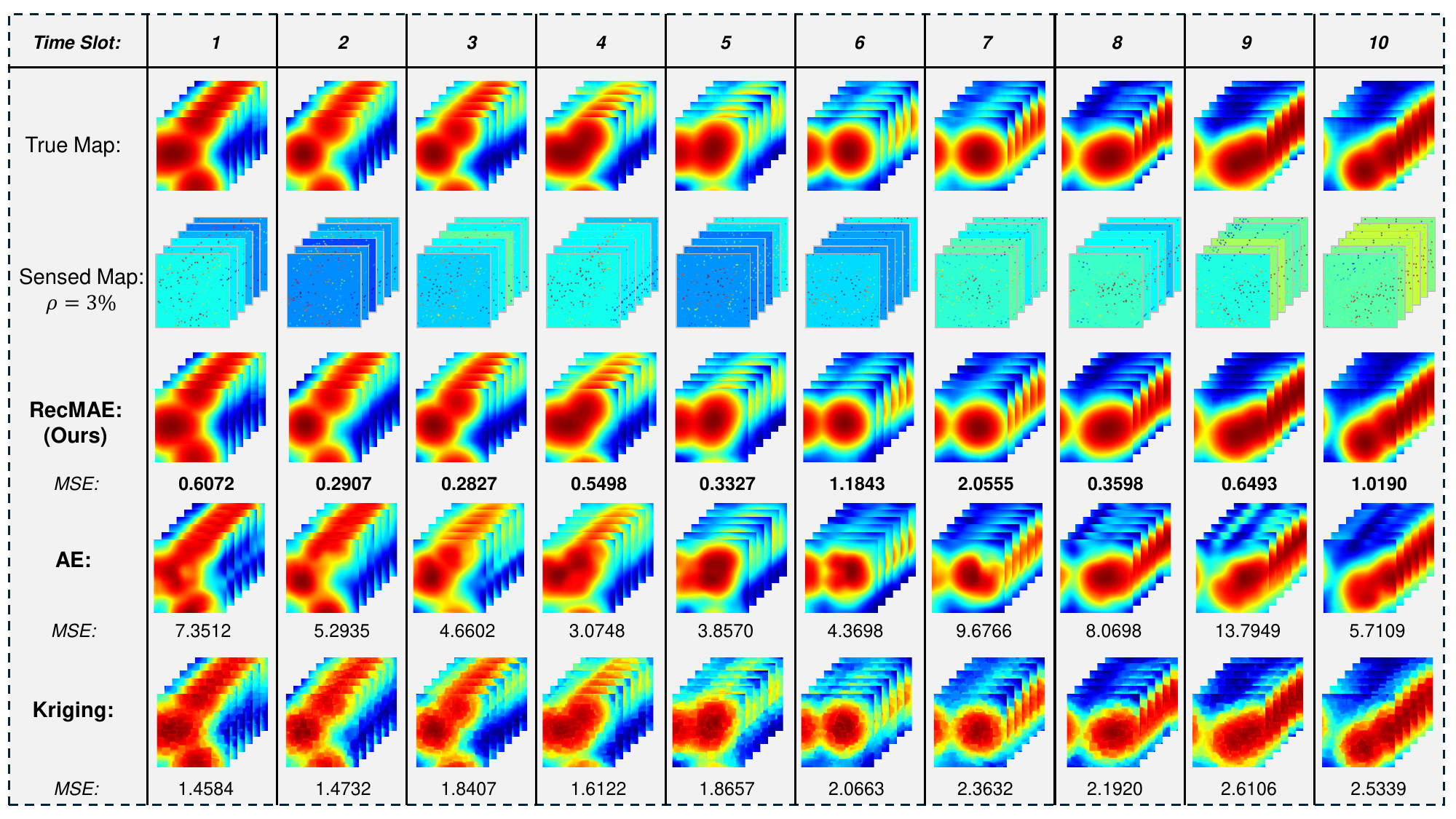}
    \caption{
The reconstruction comparison of different methods within 10-time slots and sensing radio $\rho=3\%$.
}
\label{fig:ex1}
\end{figure*}

Fig. \ref{fig:ex1} illustrates the reconstruction error over 10 time slots at an extremely sparse coverage $\rho=3\%$ for different methods. 
From the visual comparison of the reconstructed maps, the proposed method yields results that closely resemble the original image, effectively preserving both global structure and fine-grained details. In contrast, the AE method introduces noticeable MSEs, particularly in regions with strong and weak signal variations, resulting in significant deviations. 
While the Kriging interpolation algorithm can roughly localize signal strengths, which achieves a lower MSE than the AE baseline by approximately 60\%, its performance is heavily dependent on the choice of the kernel function. It often leads to uneven interpolation, compromising the quality of the reconstructed map.
These substantial improvements underscore the effectiveness of our approach for map reconstruction under the same sensing conditions.

\subsection{Performance Analysis of the GenAI Planning Stage}

To evaluate the performance of the proposed MADP, 
we consider a temporal spectrum reconstruction scenario with $n_T = 10$ discrete time slots. In each round of spectrum reconstruction, the UAV starts from a fixed initial state and infers the distance of the next move based on the currently acquired perception information.

\textbf{Baseline Methods:}
We compare the proposed MADP framework with several baseline methods, including several different DRL models. The key baseline methods and their abbreviations are as follows:
\begin{itemize}
    \item[-] \textbf{CNN}: The standard MADDPG algorithm, where CNNs are used for feature extraction \cite{lowe2017multi}. 
    \item[-] \textbf{CNN-Attention}: An enhanced MADDPG variant that incorporates the temporal-attention state encoder described in Section~\ref{sec:tattn}.
    \item[-] \textbf{Random}: A baseline where the UAVs' movements are selected randomly without any policy guidance.
\end{itemize}

\textbf{Experiment Settings:}
In this section, we consider the same scenario setting as in the previous experiments, where each grid cell has a size of $4\,\mathrm{m} \times 4\,\mathrm{m}$.
We assume that dynamic sensors can sense a surrounding $3 \times 3$ grid of radio signals, whereas static sensors are limited to sensing only a $1 \times 1$ area \cite{shrestha2022spectrum}. The static sensors are assumed to be evenly distributed across the environment. 
During each time step, the dynamic sensors can move up to two grid cells in either the east-west or north-south direction.
The learning models are trained over 2000 epochs. During the initial 500 epochs, the proportion of random exploration is gradually reduced. From epoch 1300 onward, the learning rate progressively decreases until it reaches zero.
The reward function is defined as $30$ minus the error to ensure reward distribution remains unbiased.
For the proposed MADP algorithm, we set the number of denoising steps $T$ as 6.

\begin{figure}[htbp]
\centering
\begin{subfigure}{.28\textwidth}
  \centering
  \includegraphics[width=1.0\linewidth]{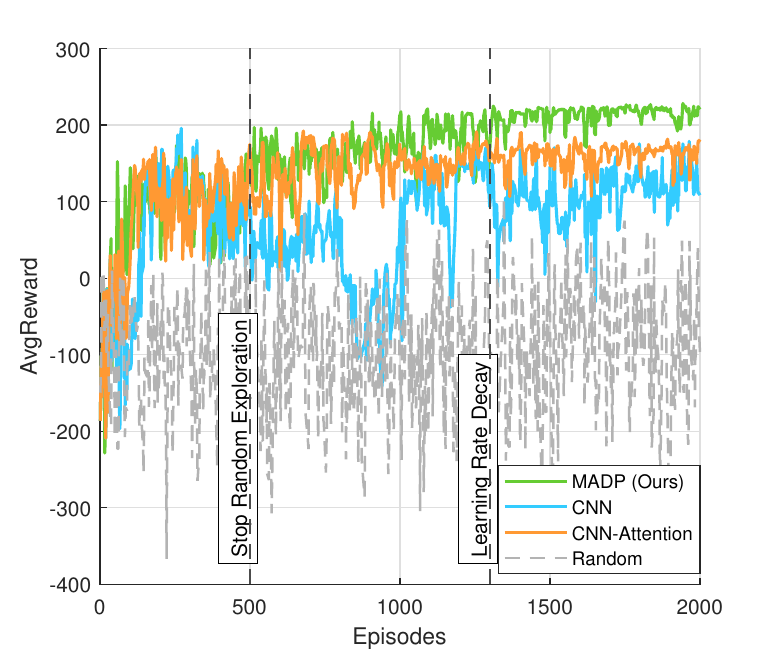} 
  \caption{Training Curve}
  \label{fig:ex2_1_sub1}
\end{subfigure}%
\begin{subfigure}{.21\textwidth}
  \centering
  \includegraphics[width=1.0\linewidth]{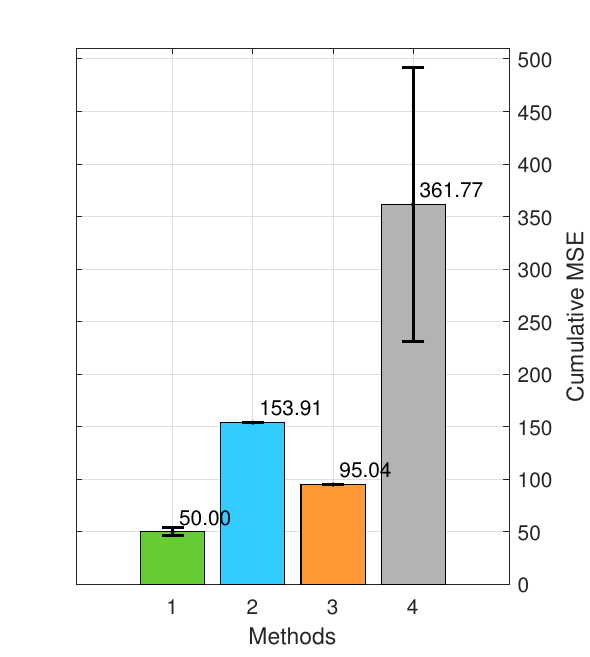}
  \caption{Cumulative MSE}
  \label{fig:ex2_1_sub2}
\end{subfigure}
\caption{The training curve and cumulative MSE of 4 methods}
\label{fig:ex2_1}
\end{figure}

\textbf{Evaluation:}
First, we evaluate the performance of the proposed MADP in the GenAI planning stage by comparing it against baseline strategies and analyzing its learning behavior. 
Fig. \ref{fig:ex2_1} presents the learning curves and final reconstruction errors for MADP and three baseline methods under a given scenario, where the number of UAVs is 4 and static sensor spacing is 16.

As shown in Fig. \ref{fig:ex2_1_sub1}, MADP achieves a substantially higher average reward during training, converging faster and more stably than the alternatives. 
The MADP agents’ reward generally improves over time, showing quick recovery from occasional drops, particularly as random exploration is gradually reduced after 500 epochs and the learning rate decays after about 1300 epochs. In contrast, the CNN and CNN-Attention baselines improve more slowly, with CNN-Attention exhibiting larger fluctuations in reward, even when the learning rate is low.
By the end of the training, MADP attains the highest reward and demonstrates the most stable learning trajectory, indicating superior training stability.
Fig. \ref{fig:ex2_1_sub2} compares the cumulative reconstruction MSE achieved by each method during the execution process, averaged over five repeated experiments.
The proposed MADP achieves the lowest MSE of 50.00, significantly outperforming both the CNN baseline at 153.91 and the CNN-Attention variant at 95.04.
In other words, MADP reduces the overall cumulative MSE by about 67.51\% relative to the standard CNN-based MADDPG and about 47.39\% relative to the attention-augmented baseline. The Random policy is the worst, with an MSE of 361.55, roughly triple the MSE of MADP.
These results highlight the effectiveness of our diffusion-based multi-agent planner in minimizing reconstruction error. Notably, incorporating temporal attention does improve the baseline over the plain CNN, confirming that capturing temporal context aids learning.
The lowest cumulative reconstruction MSE indicates improved learning stability, resulting from the generative policy’s more effective exploration-exploitation balance.

\begin{figure}[htbp]
\centering
\begin{subfigure}{.28\textwidth}
  \centering
  \includegraphics[width=1.0\linewidth]{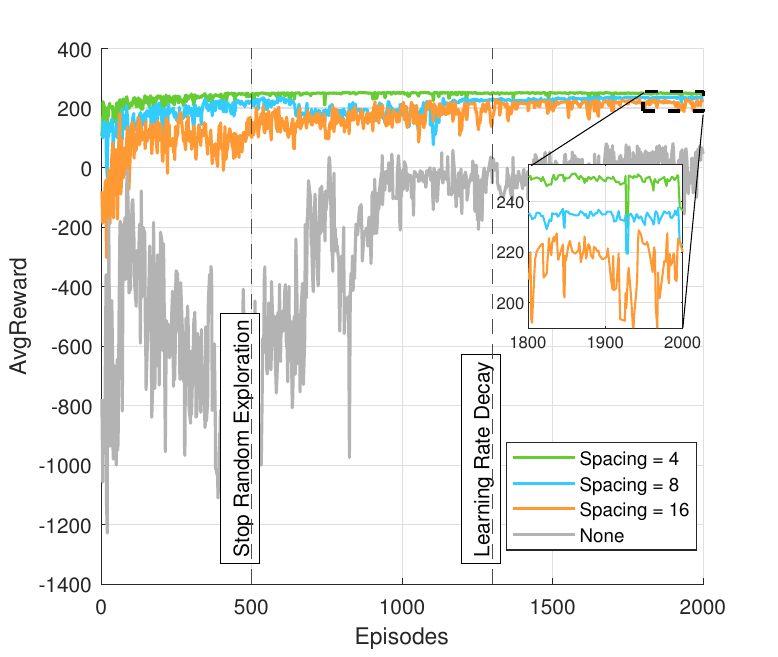} 
  \caption{Training Curve}
  \label{fig:ex2_2_sub1}
\end{subfigure}%
\begin{subfigure}{.21\textwidth}
  \centering
  \includegraphics[width=1.0\linewidth]{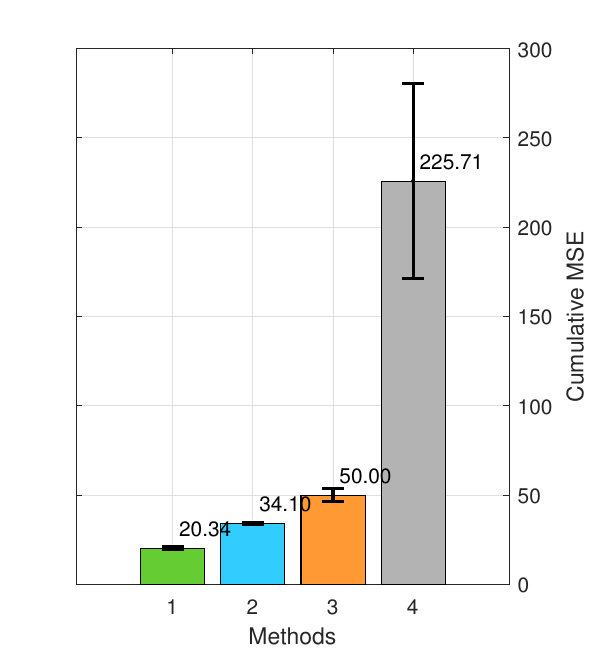}
  \caption{Cumulative MSE}
  \label{fig:ex2_2_sub2}
\end{subfigure}
\caption{The training curve and cumulative MSE of different spacing}
\label{fig:ex2_2}
\end{figure}

We further investigate how sensor deployment density affects performance. Fig. \ref{fig:ex2_2} examines the impact of varying the static sensor spacing on mapping accuracy, with the number of UAVs fixed at 4. As expected, having more densely distributed static sensors markedly improves reconstruction quality.
In Fig. \ref{fig:ex2_2_sub1},  the cumulative MSE achieved by MADP increases from 20.34 with a very dense static network (spacing = 4) to 34.10 at moderate density (spacing = 8) and 50.00 at sparse deployment (spacing = 16).
In the extreme case with no static sensors, the MSE spikes to 225.71
, reflecting the much heavier burden on the UAVs to sense the entire area.
Correspondingly, the training curves in Fig. \ref{fig:ex2_2_sub2} show faster convergence and higher final rewards when more static measurements are available, since the GenAI reconstructor can rely on richer initial reducing requirements for dynamic sensor locations.
When static sensors are removed, the MADP agents can still learn a policy to cover the area; however, convergence is slower, and the final reward is lower due to increased uncertainty. These trends indicate that while our MADP framework can function with only mobile sensors, having even a sparsely distributed static sensor grid significantly enhances mapping performance by providing valuable observations.

\begin{figure}[htbp]
\centering
\begin{subfigure}{.28\textwidth}
  \centering
  \includegraphics[width=1.0\linewidth]{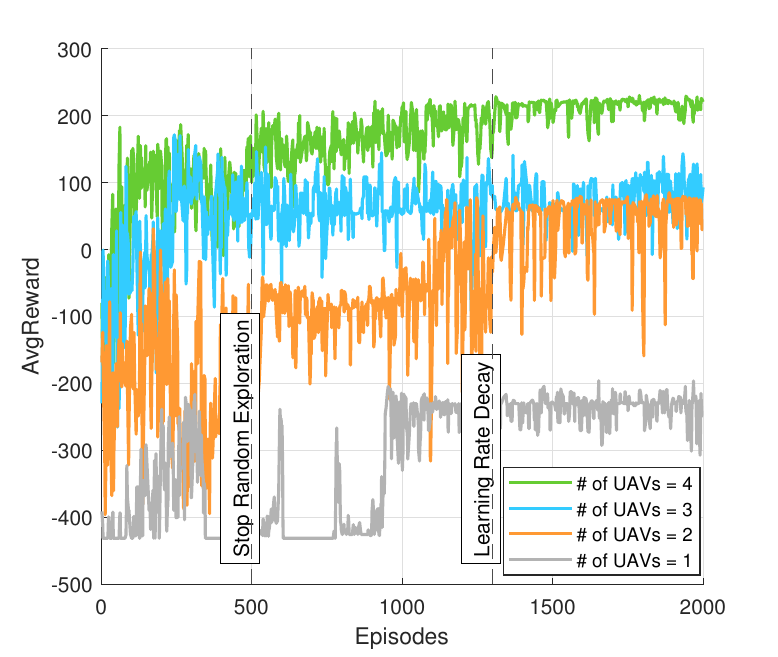} 
  \caption{Training Curve}
  \label{fig:ex2_3_sub1}
\end{subfigure}%
\begin{subfigure}{.21\textwidth}
  \centering
  \includegraphics[width=1.0\linewidth]{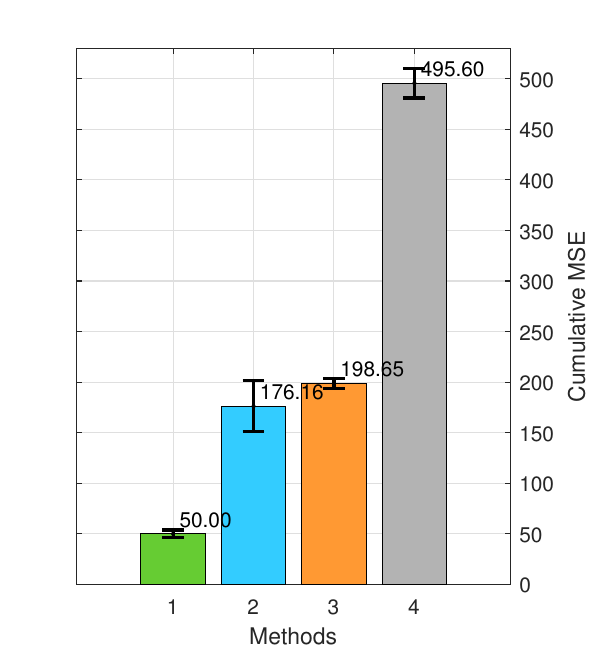}
  \caption{Cumulative MSE}
  \label{fig:ex2_3_sub2}
\end{subfigure}
\caption{The training curve and cumulative MSE of different numbers}
\label{fig:ex2_3}
\end{figure}

Fig. \ref{fig:ex2_3} illustrates the impact of UAV team size on learning performance, under the condition of a fixed static sensor spacing of 16. The results demonstrate that increasing the number of UAV agents enhances both training efficiency and final reconstruction accuracy, highlighting the benefits of multi-agent cooperation.
Particularly, using four UAVs yields a cumulative MSE of 50.00, whereas reducing to three, two, or one UAV degrades the accuracy to roughly 176.16, 198.65, and 495.60, respectively
This dramatic increase in MSE with fewer agents indicates that additional UAVs provide complementary coverage and more data, which directly translates to higher mapping fidelity. 
Unlike static sensors, UAVs exert a greater influence on reconstruction performance due to their mobility and dynamic sensing capabilities.
These observations underscore the importance of cooperative multi-UAV exploration: a greater number of agents can divide the sensing task and explore the region in parallel, thereby reducing the cumulative error and improving the robustness of the cartography process.

\begin{figure*}[htbp]
    \centering
    \includegraphics[width= 0.80\linewidth]{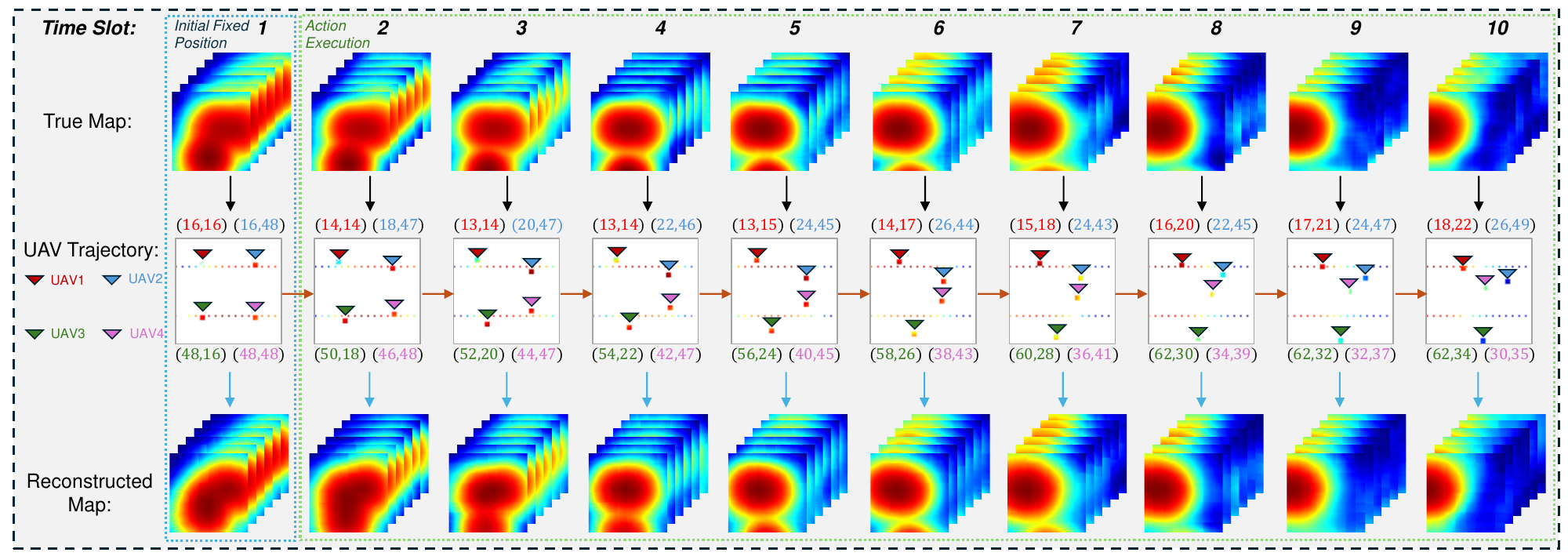}
    \caption{
The illustration of the MADP execution process and a UAV trajectory example.
}
\label{fig:ex2}
\end{figure*}

Finally, Fig. \ref{fig:ex2} illustrates the trajectories followed by the UAVs under the learned MADP policy, alongside the true spectrum map and the reconstructed map for the region.
For reconstructed maps, during the first time slot, the UAV's initial position is fixed, resulting in only partial recovery. However, starting from the second time slot, the MADP strategy gradually enhances reconstruction performance by adjusting UAVs' trajectories. Over the 10-time-slot horizon, the UAV's flight paths reflect an increasingly efficient coverage strategy.
Each UAV agent disperses to cover a different sub-area of the field, and their paths collectively ensure that diverse locations are observed over time.
Notably, the agents coordinate implicitly to avoid redundant coverage via sequentially visiting distinct waypoints.
This effective spatial exploration is evident from the correspondence between features in the true map and the reconstructed map: the learned trajectories allow the GenAI reconstructor to capture the major signal variations across the region. 
Thus, the proposed approach minimizes reconstruction error quantitatively and produces qualitatively sensible flight paths that enhance mapping performance by gathering information from all critical areas of the LAENets.

\section{Conclusion}\label{sec:sec6}

In this work, we have proposed a two-stage GenAI framework for temporal spectrum cartography in LAENets. The framework consists of a GenAI reconstructor and a GenAI planner, each responsible for one stage of the process. In the reconstruction stage, we introduced RecMAE, a masked autoencoder designed to recover temporal spectrum maps using a dual-mask mechanism. This design enhances the model’s ability to capture fine-grained details and enables more accurate spectrum reconstruction.
In the planning stage, we presented MADP, a multi-agent diffusion policy learner built upon a temporal-attention state encoder. This encoder effectively extracts temporal context from sequential observations, facilitating robust decision-making in dynamic environments.
Extensive simulation results demonstrate that our framework outperforms existing spectrum editing methods in both reconstruction and planning, enabling more effective and accurate spectrum cartography for low-altitude economy activities.

\bibliography{Ref}

\end{document}